\documentclass[sigconf,nonacm]{acmart}

\usepackage{multirow}   

\usepackage{placeins}
\usepackage{lipsum}
\usepackage{mathtools}
\usepackage{algorithm}
\usepackage{algpseudocode}

\usepackage{graphicx}      
\usepackage{booktabs}     
\usepackage{xcolor}       
\usepackage{subcaption}
\usepackage{caption} 

\usepackage[capitalize,noabbrev]{cleveref}

\usepackage{caption}

\usepackage{amsmath,amsthm}

\theoremstyle{plain}
\newtheorem{theorem}{Theorem}[section]

\theoremstyle{definition}

\theoremstyle{remark}

\usepackage[textsize=tiny]{todonotes}

\usepackage{xspace}

\AtBeginDocument{%
  }

\begin{document}

\title{SIVF: GPU-Resident IVF Index for Streaming Vector Search}

\author{Dongfang Zhao}
\email{dzhao@cs.washington.edu}
\affiliation{%
  \institution{Tacoma School of Engineering and Computing}
  \institution{Paul G. Allen School of Computer Science \& Engineering}
  \city{University of Washington}
  \country{USA}
}

\begin{abstract}
GPU-accelerated Inverted File (IVF) index is one of the industry standards for large-scale vector search but relies on static VRAM layouts that hinder real-time mutability. Our benchmark and analysis reveal that existing designs of GPU IVF necessitate expensive CPU-GPU data transfers for index updates, causing system latency to spike from milliseconds to seconds in streaming scenarios. We present SIVF, a GPU-native index that enables high-velocity, in-place mutation via a series of new data structures and algorithms, such as conflict-free slab allocation and coalesced search on non-contiguous memory. SIVF has been implemented and integrated into the open-source vector search library, Faiss. Evaluation against baselines with diverse vector datasets demonstrates that SIVF reduces deletion latency by orders of magnitude compared to the state-of-the-arts. Furthermore, distributed experiments on a 12-GPU cluster demonstrate that SIVF exhibits near perfect linear scalability, achieving an aggregate ingestion throughput of 4.07 million vectors/s and a deletion throughput of 108.5 million vectors/s.
\end{abstract}
\maketitle

\ccsdesc[500]{Information systems~Top-k retrieval in databases}

\section{Introduction}


Real-world vector analytics applications, ranging from search engines to dynamic retrieval-augmented generation (RAG) of large language models (LLMs), increasingly operate on streaming data where timeliness is critical~\cite{zli_sigmod26,jmohoney_osdi25}. These systems necessitate a sliding-window model where expired vectors must be invalidated promptly as new vectors arrive to maintain bounded memory footprint. However, existing GPU-accelerated approximate nearest neighbor (ANN) indices are predominantly optimized for static, write-once-read-many workloads~\cite{zzhang_nsdi24,vectraflow}.

To quantify the insertion and eviction performance, we conducted benchmarks with the industry-standard Faiss library~\cite{johnson2019billion} and the state-of-the-art graph index CAGRA (cuVS)~\cite{10597683} on an NVIDIA RTX 6000 GPU hosted at the Chameleon Cloud~\cite{keahey2020lessons}. We utilized the SIFT1M dataset~\cite{jegou2011product} to simulate a realistic sliding-window scenario, measuring the wall-clock latency of ingesting 10,000 new vectors and evicting the oldest 10,000.

\begin{figure}[!t]
    \centering
    \begin{subfigure}[b]{0.49\linewidth}
        \centering
        \includegraphics[width=\linewidth]{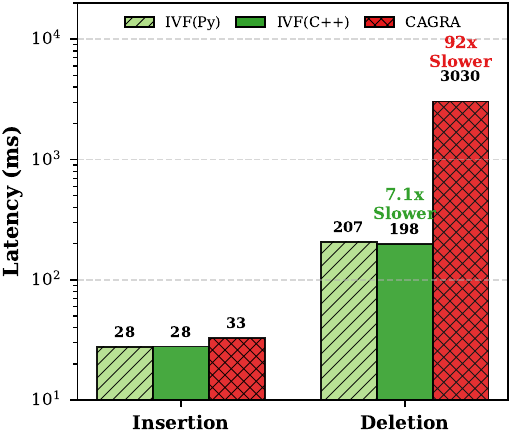}
        \caption{Physical deletion overhead.}
        \label{fig:motivation_physical}
    \end{subfigure}
    \hfill
    \begin{subfigure}[b]{0.49\linewidth}
        \centering
        \includegraphics[width=\linewidth]{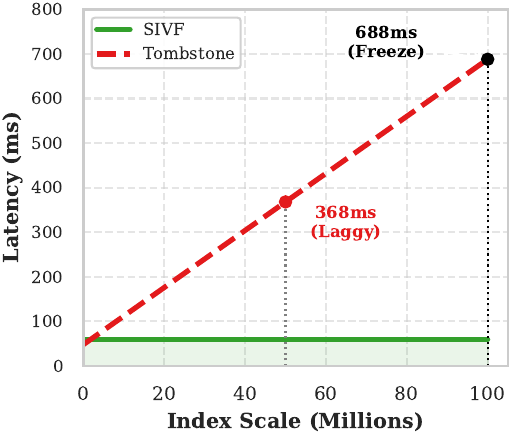}
        \caption{Tombstone scalability trap.}
        \label{fig:motivation_tombstone}
    \end{subfigure}
    
    \caption{The dilemma of mutability on GPUs.}
    \Description{The dilemma of mutability on GPUs. (a) Insertion is accelerated by GPU parallelism, but deletion incurs a prohibitive CPU-GPU roundtrip. (b) Tombstone deletion avoids data movement but leads to unbounded memory growth and degraded query performance.}
    \label{fig:motivation}
\end{figure}

As illustrated in Figure~\ref{fig:motivation}(a), we observe a drastic asymmetry between insertion and physical deletion. While GPU parallelism accelerates insertion to $\sim$28 ms for IVF, the eviction process is significantly slower. For Faiss IVF, physical deletion spikes the latency to $\sim$200 ms ($\sim$7.1$\times$ slower) due to the contiguous memory layout necessitating expensive data shifting. The situation is catastrophic for graph-based indices: CAGRA incurs a deletion latency of over 3,000 ms ($\sim$92$\times$ slower) as the topology requires reconstruction to maintain connectivity of the underlying graph.

A common alternative is \textit{lazy deletion} (aka tombstone), which marks vectors as invalid to avoid immediate physical movement. However, this strategy merely defers the cost. As shown in Figure~\ref{fig:motivation}(b), tombstone mechanisms suffer from poor scalability due to the $O(N)$ complexity of garbage collection (aka compaction). Based on our microbenchmarks, the compaction pause scales linearly with index size. While negligible at small scales, the latency is projected to exceed 360 ms at 50 million vectors and reach nearly 700 ms at 100 million vectors. These results confirm a fundamental limitation: current GPU indices lack an efficient mutability mechanism that is both low-latency (unlike physical deletion) and scalable (unlike tombstones), necessitating a new indexing method ideally with a constant cost, as represented by SIVF that this paper will present.

Our analysis of the Faiss source code~\cite{douze2024faiss} further reveals that this performance asymmetry stems from a rigid class hierarchy that enforces a fallback to host-side processing on CPUs. In the library's architecture, the data eviction interface is defined in the abstract base class \texttt{faiss::Index} via the \texttt{remove\_ids} virtual method. While CPU-based implementations override this method to provide efficient in-memory deletion logic (e.g., \texttt{memmove}), the GPU counterparts inherit the base implementation which lacks native support. Consequently, current GPU indices rely on a \emph{CPU-GPU Roundtrip} pattern for eviction. That is, to delete even a single vector, the system must transfer the entire index state from VRAM to host memory, perform the compaction on the CPU, and re-upload the modified index to the device. This heavy I/O burden saturates the PCIe bus and prevents the system from utilizing the GPU's compute throughput, capping the maximum sustainable I/O rate in streaming scenarios.


The lack of a GPU-resident eviction operator is not an oversight, but a consequence of how GPU IVF indices are engineered for throughput. To maximize memory coalescing and bandwidth utilization, inverted lists are typically stored in pre-allocated contiguous buffers. In this setting, naive tombstone deletion is hard to make practical. First, tombstones provide only logical deletion and do not reclaim space, so under a sliding-window workload dead entries accumulate and lists grow, forcing queries to scan and filter more invalid slots unless compaction is performed. Second, compaction and slot reuse in VRAM require bulk data movement and careful synchronization with concurrent queries and updates under the GPU memory model. 
In addition, standard InVerted File (IVF) indices lack an ID-to-Location mapping. Without a reverse index, locating a specific ID for deletion requires scanning all inverted lists, an operation with $O(N)$ complexity that negates the benefits of GPU acceleration. These constraints make naive GPU eviction prohibitively inefficient.



To resolve these limitations, we propose SIVF, a GPU-resident indexing method that supports high-throughput mutability through four specialized device-side algorithms. First, to enable nonblocking writes, we design a \emph{Lock-Free Ingestion} protocol (Algorithms~\ref{alg:sdma_core} \& \ref{alg:sivf_ingestion}) that manages a pool of fixed-size slabs. This protocol utilizes \texttt{atomic Compare-And-Swap} for slot reservation and speculative head updates, allowing concurrent threads to append to linked lists without global locks. Second, to guarantee data consistency under CUDA's relaxed memory model, we apply a strict publication ordering; the kernel issues device-wide memory fences (\verb|__threadfence()|) before committing validity bits, ensuring readers never observe partially initialized payloads. Third, to mitigate the latency of pointer chasing, we propose \emph{Warp-Cooperative Search} (Algorithm~\ref{alg:sivf_search}). By aligning the slab capacity ($C=32$) with the hardware warp width, this algorithm enables threads to cooperatively load and evaluate non-contiguous memory blocks while preserving coalesced access patterns. Finally, we address deletion by maintaining a GPU-resident Address Translation Table along with lazy eviction (Algorithm~\ref{alg:sivf_delete}), allowing SIVF to decouple logical removal from physical data movement, which makes eviction to a constant-time operation.

SIVF has been implemented and integrated into Faiss~\cite{douze2024faiss} as a new index through the \texttt{GpuIndexSIVF} class. Our implementation leverages CUDA for device-side kernels and employs a dual-view memory management strategy to maintain the consistency between host and device states. All new algorithms as well as key data structures are encapsulated within a \texttt{SlabManager} class that abstracts the complexity of dynamic memory operations on the GPU. 

We have evaluated SIVF against a comprehensive set of baselines within Faiss, including CAGRA (cuVS)~\cite{10597683}, IVF~\cite{johnson2019billion}, Flat~\cite{douze2024faiss}, HNSW~\cite{malkov2018efficient}, NSG~\cite{10.14778/3303753.3303754}, and LSH~\cite{10.1145/997817.997857}. Our evaluation was carried out on six datasets representing a variety of modalities, dimensions, and scalability: synthetic~\cite{douze2024faiss}, Deep1B~\cite{babenko2016efficient}, SIFT1M~\cite{jegou2011product}, T2I-1B~\cite{simhadri2025results}, GIST1M~\cite{jegou2011product}, and DINO10B~\cite{9709990}. Experimental results show that SIVF achieves up to 1765x reduction in deletion latency compared to CPU fallback baselines and improves ingestion throughput by 36x to 120x compared to existing GPU indices. Furthermore, distributed evaluations on a 12-GPU cluster demonstrate that SIVF exhibits near-perfect linear scalability, reaching an aggregate deletion throughput of 108.5 million vectors/s and 159.3K vectors/s for high-fidelity search. In end-to-end sliding window scenarios, SIVF delivers a 163x to 262x speedup with a negligible memory overhead less than 0.8 percent, effectively closing the gap between static and streaming vector analytics performance.

In summary, this paper makes the following contributions:
\begin{itemize}
    \item We identify the CPU-GPU bottleneck in existing GPU vector indices when being used for sliding-window scenarios, demonstrating that the lack of native deletion support renders them unsuitable for streaming vector analytics.
    
    \item We propose a new indexing method, namely SIVF, that synthesizes a series of GPU-optimized data structures and algorithms for streaming vector analytics. This design enables $O(1)$ time complexity for deletion and constant-time insertion overhead with negligible memory overhead.
    
    \item We implement SIVF in the state-of-the-art vector search library Faiss and evaluate SIVF against a broad spectrum of baselines including GPU CAGRA, GPU Flat, GPU IVF, HNSW, NSG, and LSH on 6 representative datasets and a 12-GPU cluster. The results demonstrate that SIVF achieves up to 1765x reduction in deletion latency, up to 120x faster ingestion compared to existing GPU indices, and exhibits near perfect linear scalability in distributed environments reaching 108.5 million vectors/s for deletion and 4.07 million vectors/s for ingestion on 12 GPUs. These advancements transform vector search from a static retrieval task into an integral component of real-time vector data analytics.
\end{itemize}



\paragraph{Paper Organization}
The remainder of this paper is organized as follows. 
Section~\ref{sec:related_work} reviews the existing landscape of vector search, distinguishing between CPU-based dynamic methods and static GPU-accelerated indices. 
Section~\ref{sec:design_implementation} details the architectural design of SIVF, elaborating on the slab-based memory allocator, the validity bitmap mechanism, and the GPU-resident address translation table. 
Section~\ref{sec:implementation_details} describes the implementation of SIVF within the Faiss library, including key data structures and kernel optimizations.
Section~\ref{sec:evaluation} presents our experimental methodology and a comprehensive evaluation of SIVF against state-of-the-art baselines. 
Finally, Section~\ref{sec:conclusion} concludes the paper and outlines directions for future research.

\section{Related Work}
\label{sec:related_work}


\subsection{Optimization of Graph-Based Indices}

Graph-based indices remain the state-of-the-art for high-recall retrieval. To address performance bottlenecks in production, frameworks like VSAG~\cite{10.14778/3750601.3750624} optimize memory access patterns and parameter tuning, while SOAR~\cite{DBLP:conf/nips/SunSDGK23} improves indexing structures. 
Of note, CAGRA~\cite{10597683} establishes a new state-of-the-art for static graph construction and search throughput on GPUs; however, it is fundamentally designed for write-once-read-many workloads and lacks native support for the efficient, fine-grained in-place deletions required by streaming applications.
Navigational efficiency is a key optimization target; SHG~\cite{10.14778/3748191.3748212} introduces a hierarchical graph with shortcuts to bypass redundant levels, and probabilistic routing methods~\cite{lu2024probabilistic} have been proposed to enhance traversal. 
Several works focus on distance metric optimizations: ADA-NNS~\cite{10.1145/3696410.3714870} utilizes angular distance guidance to filter irrelevant neighbors, DADE~\cite{10.14778/3712221.3712244} accelerates comparisons via data-aware distance estimation in lower dimensions, and HSCG~\cite{10.1145/3769786} adapts the Monotonic Relative Neighbor Graph for cosine similarity using hemi-sphere centroids.
Efforts to improve graph construction and maintenance include LIGS~\cite{10.1145/3726302.3730028}, which employs locality-sensitive hashing (LSH) to simulate proximity graphs for faster updates, and CSPG~\cite{DBLP:conf/nips/YangCZ24}, which crosses sparse proximity graphs. 
Addressing robustness, Hua et al.~\cite{10.1145/3769783} propose dynamically detecting and fixing graph hardness for out-of-distribution queries. MIRAGE-ANNS~\cite{10.1145/3725325} attempts to bridge the gap between incremental and refinement-based construction. 
Furthermore, theoretical frameworks like Subspace Collision~\cite{10.1145/3709729} provide guarantees on result quality via clustering-based indexing.

Collectively, while these approaches maximize navigational efficiency for static data, they fundamentally rely on complex edge dependencies that are prohibitively expensive to maintain on GPUs. Unlike SIVF, they lack the architectural support for $O(1)$ lock-free mutation, rendering them unsuitable for high-churn streaming workloads where sub-millisecond update latency is critical.

\subsection{Attribute-Filtered and Hybrid Search}

The integration of structured constraints with vector analytics has led to a surge in Filtered ANN research. Li et al.~\cite{10.1145/3769763} provide a comprehensive taxonomy and benchmark of these methods. 
For range filtering, UNIFY~\cite{10.14778/3717755.3717770} introduces a segmented inclusive graph to support pre-, post-, and hybrid filtering strategies. 
Dynamic environments are addressed by DIGRA~\cite{10.1145/3725399}, which uses a multi-way tree structure for dynamic graph indexing, and RangePQ~\cite{10.1145/3725401}, which offers a linear-space indexing scheme for range-filtered updates. 
Peng et al.~\cite{10.14778/3748191.3748193} propose dynamic segment graphs to handle mixed streams of data and queries.
Regarding specific constraints, Wang et al.~\cite{DBLP:conf/nips/WangLX0YN23} present a robust framework for general attribute constraints. 
Engels et al.~\cite{engels2024approximate} focus on window filters, while WoW~\cite{10.1145/3769843} develops a window-graph based index to handle window-to-window incremental construction.

However, these contributions primarily optimize algorithmic filtering logic rather than the underlying storage architecture. They remain constrained by the static memory layouts of standard GPU indices, whereas SIVF fundamentally redesigns the VRAM management to enable the high-throughput, in-place mutability that is a prerequisite for such dynamic operations.

\subsection{Quantization and Scalability}

To manage high-dimensional data at scale, quantization and system-level optimizations are critical. 
RaBitQ~\cite{10.1145/3654970} and its extended version~\cite{10.1145/3725413} provide theoretical error bounds for bitwise quantization with flexible compression rates. 
SymphonyQG~\cite{10.1145/3709730} integrates quantization codes directly with graph structures to reduce memory access overhead, while LoRANN~\cite{DBLP:conf/nips/JaasaariHR24} utilizes low-rank matrix factorization for compression.
In terms of system architecture, Tagore~\cite{10.1145/3769825} leverages GPUs for accelerating graph index construction. 
For distributed environments, HARMONY~\cite{10.1145/3749167} employs a multi-granularity partition strategy to balance load. 
WebANNS~\cite{10.1145/3726302.3730115} enables efficient search within web browsers via WebAssembly. 
From a theoretical perspective, Indyk and Xu~\cite{DBLP:conf/nips/IndykX23} analyze the worst-case performance limits of popular implementations. 
Finally, DARTH~\cite{10.1145/3749160} introduces declarative recall targets through adaptive early termination to balance performance and result quality.

While these techniques reduce storage footprints or distribute load, they predominantly operate on static memory layouts that are brittle under frequent updates. SIVF addresses a complementary challenge: it provides the dynamic memory substrate that allows these scalable architectures to support real-time streams without succumbing to VRAM fragmentation or locking overheads.

\subsection{GPU Acceleration}
Recent HPC work highlights how careful kernel and data-layout design can unlock GPU performance across data-intensive workloads. FZ-GPU demonstrates a fully parallel compression pipeline with both warp-aware bitwise operations and shared-memory efficiency~\cite{10.1145/3588195.3592994}. For sparse linear algebra, SpMV designs reduce preprocessing overhead while improving load balance and memory access locality~\cite{10.1145/3588195.3593002}, and GPU-aware preconditioning further emphasizes locality and coalescence as first-order concerns on modern GPU architectures~\cite{10.1145/3625549.3658683}. Beyond compute kernels, GPU-enabled asynchronous checkpoint caching and prefetching treat GPU memory as a first-class tier to improve end-to-end throughput for I/O-heavy workflows~\cite{10.1145/3588195.3592987}. Format choices and composition are also critical for irregular workloads, as shown by automatic sparse format composition for SpMM that avoids expensive autotuning while delivering strong performance~\cite{10.1145/3731545.3731574}.
Another line of work focuses on reliability, debugging, and cluster-scale GPU management, which together shape how GPU-centric systems are built and operated. GPU-FPX provides low-overhead floating-point exception tracking for NVIDIA GPUs~\cite{10.1145/3588195.3592991}, FPBOXer improves scalable input generation for triggering floating-point exceptions in GPU programs~\cite{10.1145/3625549.3658660}, and FloatGuard extends whole-program exception detection to AMD GPUs and highlights cross-vendor portability concerns~\cite{10.1145/3731545.3731586}. At the application and cluster level, SIMCoV-GPU studies multi-node, multi-GPU acceleration for irregular agent-based simulations~\cite{10.1145/3625549.3658692}, FASOP automates fast search for near-optimal transformer parallelization on heterogeneous GPU clusters~\cite{10.1145/3625549.3658687}, and serverless platforms motivate GPU sharing and scheduling mechanisms such as ESG and FluidFaaS~\cite{10.1145/3625549.3658657,10.1145/3731545.3731580}.

However, applying these general-purpose HPC optimizations to vector search remains non-trivial due to the irregular memory access patterns of IVF traversal. SIVF bridges this gap by tailoring these micro-architectural principles, specifically warp-aligned slabs and lock-free synchronization, to construct the first GPU-resident index capable of sustaining high-velocity updates without compromising retrieval integrity.

\section{SIVF: Streaming IVF Indexing for GPUs}
\label{sec:design_implementation}



This section presents the design of SIVF for high-concurrency updates on GPUs. We first introduce the Slab-based Dynamic
Memory Allocator (SDMA) in Section~\ref{sec:slab_memory_management}, which represents each IVF list as a linked chain of fixed-capacity
slabs with per-list head pointers $H[\cdot]$ and per-slab metadata 
\[(\texttt{next}, \texttt{valid\_count}, \texttt{validity\_bitmap}),\]
summarized in Algorithm~\ref{alg:sdma_core}. Building on SDMA, we describe the parallel ingestion (Section~\ref{sec:lock_free_ingestion})
in which each thread inserts one vector by reserving a slot via \texttt{atomicCAS} on \texttt{valid\_count} and publishing visibility
by setting a validity bit after \verb|__threadfence()| (Algorithm~\ref{alg:sivf_ingestion}). We then present a warp-cooperative search (Section~\ref{sec:methods_search_kernel}) that assigns one warp to one query and evaluates one slab per traversal step by
consulting the validity bitmap before reading payloads (Algorithm~\ref{alg:sivf_search}). We detail lazy eviction
(Section~\ref{sec:bitmap_lazy_eviction}), which performs constant-time deletion by looking up coordinates in the address table
$\mathcal{T}$ and atomically clearing the corresponding bitmap bit (Algorithm~\ref{alg:sivf_delete}).
We demonstrate the theoretical properties of SIVF in Section~\ref{sec:theoretical_analysis}.

\paragraph{Data Model and Assumptions}
To ensure $O(1)$ address resolution without hashing collisions, SIVF operates on a dense identifier space $v_{id} \in [0, N_{max})$, where $N_{max}$ is the pre-configured capacity. In production environments, sparse or non-integer external IDs (e.g., 64-bit UUIDs) are mapped to this internal dense range via a lightweight host-side registry. Furthermore, SIVF enforces strong consistency for data updates via a ``delete-then-insert'' semantic: overwriting an existing $v_{id}$ involves atomically clearing its existing validity bit (logical deletion) before the new payload becomes visible. This design ensures that the validity bitmap serves as the single source of truth for searchers.

\subsection{Slab-based Memory Management}
\label{sec:slab_memory_management}

We propose a Slab-based Dynamic Memory Allocator (SDMA) that replaces monolithic, variable-length inverted lists with linked chains of fixed-size blocks on GPU. SDMA pre-allocates a contiguous slab pool; each slab stores up to $C$ vectors and a lightweight metadata header $M=\langle n_{next}, b_{valid}, \text{cnt} \rangle$, where $n_{next}$ is the next-slab pointer, $b_{valid}$ is a $C$-bit validity bitmap, and $\text{cnt}$ tracks the number of active vectors. We set $C=32$ to match the GPU warp size. Each IVF list $\ell$ is represented by a head pointer $H[\ell]$ to the first slab in its chain. 

Algorithm~\ref{alg:sdma_core} summarizes these core primitives.
Allocation draws a fresh slab id from a global stack pointer $P_{top}$ through this primitive: $s=\textnormal{\texttt{atomicSub}}(P_{top},1)-1$, initializes its metadata, and links it to the list head $H[\ell]$ using an atomic compare-and-swap (CAS) operation. Insertion for a vector id $v_{id}$ first resolves the target slab $s$, reserves a free slot $o$, writes the payload, and finally commits visibility by atomically setting bit $o$ in $b_{valid}$.
To support $O(1)$ \textit{physical} deletion without expensive garbage collection, SDMA maintains an address table $\mathcal{T}$ mapping each id to its physical coordinate $\mathcal{T}(v_{id})=\langle s,o\rangle$. The deletion operation performs a direct lookup and atomically clears bit $o$ in $b_{valid}$. Unlike tombstone-based designs, SDMA performs immediate reclamation: it atomically decrements the slab's occupancy counter $\text{cnt}$. If $\text{cnt}$ drops to zero, the slab is instantly recycled by pushing it back to the global free stack $P_{top}$, making it available for future insertions. 

\begin{algorithm}[t]
\caption{Core operations of SDMA}
\label{alg:sdma_core}
\begin{algorithmic}[1]
\State \textbf{Global State:} slab pool metadata and data, per-list head pointers $H[\cdot]$, stack pointer $P_{\text{top}}$, address table $\mathcal{T}$
\State \textbf{Input:} vector identifier $v_{\text{id}}$, vector $x$, list assignment $\ell$

\Procedure{Insert}{$v_{\text{id}}, x, \ell$}
    \State $s \gets H[\ell]$
    \While{$s$ is invalid \textbf{or} no free slot exists in slab $s$}
        \State $s_{\text{new}} \gets \textnormal{\texttt{atomicSub}}(P_{\text{top}}, 1) - 1$
        \State Initialize metadata $s_{\text{new}}$ with $b_{\text{valid}}=0$, $n_{\text{next}}=s$
        \State Update $H[\ell]$ to $s_{\text{new}}$ via \textnormal{\texttt{atomicCAS}}
        \State $s \gets H[\ell]$
    \EndWhile
    \State Reserve a free slot $o$ in slab $s$ via atomic update of $c_{\text{valid}}$
    \State Write $x$ to slab data[$s$][$o$]
    \State Atomically set bit $o$ in $b_{\text{valid}}$
    \State $\mathcal{T}(v_{\text{id}}) \gets \langle s, o \rangle$
\EndProcedure

\Procedure{Delete}{$v_{\text{id}}$}
    \State $\langle s, o \rangle \gets \mathcal{T}(v_{\text{id}})$
    \If{$\langle s, o \rangle \neq \textnormal{\texttt{INVALID}}$}
        \State $old\_map \gets \text{Atomically clear bit } o \text{ in } b_{\text{valid}} \text{ of slab } s$
        
        \If{bit $o$ was set in $old\_map$} 
            \State $\mathcal{T}(v_{\text{id}}) \gets \textnormal{\texttt{INVALID}}$
            
            \State $old\_cnt \gets \textnormal{\texttt{atomicSub}}(\&\text{slab\_meta}[s].\text{cnt}, 1)$
            \If{$old\_cnt == 1$} 
                \State Atomically push $s$ back to $P_{\text{top}}$ stack
                \State $\text{slab\_meta}[s].b_{\text{valid}} \gets 0$
            \EndIf
        \EndIf
    \EndIf
\EndProcedure
\end{algorithmic}
\end{algorithm}

\subsection{Lock-free Parallel Ingestion}
\label{sec:lock_free_ingestion}

SIVF uses a fully parallel ingestion kernel in which each CUDA thread inserts one vector into its assigned IVF list $\ell$.
Each list is a singly linked chain of fixed-capacity slabs, with head pointer $H[\ell]$ and per-slab metadata
\texttt{valid\_count}, \texttt{validity\_bitmap}, and \texttt{next}. Insertion follows a reserve-write-publish protocol.
A thread first reads the current head slab $h=H[\ell]$ and attempts to reserve a slot by incrementing
\texttt{slab\_meta[$h$].valid\_count} using \texttt{atomicCAS}; the reserved slot index is the old counter value $c$.
On success, the thread writes the payload to \texttt{slab data[$h$][$c$]} and the id to \texttt{slab\_ids[$h$][$c$]}, updates
the address table $\mathcal{T}(v_{id})\gets\langle h,c\rangle$, then executes \verb|__threadfence()| and sets the corresponding
validity bit using \texttt{atomicOr}. The validity bit is the only publication signal read by search, so this ordering ensures
that a reader observing the bit set never sees partially initialized payload or a missing table entry. 
If the head slab is
absent or full, the thread allocates a new slab id from the global pool via \texttt{atomicSub} on $P_{top}$, initializes its
metadata (\texttt{valid\_count}, \texttt{validity\_bitmap}, \texttt{next}), executes \verb|__threadfence()|, and attempts to
publish it as the new head with \texttt{atomicCAS(\&H[$\ell$], h, s\_new)}. 
If head publication fails under contention (i.e., another thread updated the head first), the thread immediately reclaims the allocated slab to the global free list to prevent memory leaks and retries the operation.
If the slab pool is exhausted, the insertion fails fast for that element and returns an error to the caller, which can throttle the update stream or retry later, rather than silently dropping updates.

The full per-thread protocol is described in Algorithm~\ref{alg:sivf_ingestion}. Algorithm~\ref{alg:sivf_ingestion} details the concurrent ingestion protocol where each thread inserts a vector independently without acquiring global locks. The logic is divided into two main execution paths depending on the capacity of the current head slab.
(1) If the current head slab possesses available slots, the thread attempts to reserve an index by executing an atomic compare and swap operation on the occupancy counter. Winning this atomic operation guarantees exclusive write access to the specific slot. The thread then writes the vector payload and identifier, and maps the physical coordinates in the address translation table. A device memory fence is explicitly issued to guarantee that all payload writes are fully committed to memory before the slot is logically published. The vector finally becomes visible to concurrent searchers when the thread atomically sets the corresponding bit in the validity bitmap.
(2) If the head slab is completely full or does not exist, the thread fetches a fresh slab from the global pool via an atomic subtraction on the global stack pointer. The thread initializes the metadata of this new slab and links its next pointer to the old head. It then attempts to swing the list head pointer to the newly allocated slab using an atomic compare and swap. A successful swap allows the thread to populate the first slot and publish the bitmap. If the swap fails due to concurrent modifications by other threads, the thread avoids a memory leak by immediately returning the unused slab to the global free list via an atomic addition, after which it retries the entire insertion process. This retry mechanism ensures safe forward progress even under high contention.

\begin{algorithm}[t]
\caption{Lock-free ingestion protocol in SIVF}
\label{alg:sivf_ingestion}
\begin{algorithmic}[1]
\State Input: vector $x$, identifier $v_{id}$, list assignment $\ell$
\State Global: head array $H[\cdot]$, slab\_meta, slab\_data, free list, stack pointer $P_{top}$, address table $\mathcal{T}$
\State $attempts \gets 0$
\While{$attempts < \textsc{MAX\_INSERT\_ATTEMPTS}$}
    \State $attempts \gets attempts + 1$
    \State $h \gets H[\ell]$
    \If{$h \neq -1$}
        \State $c \gets \text{slab\_meta}[h].\text{valid\_count}$
        \If{$c < C$}
            \If{$\texttt{atomicCAS}(\&\text{slab\_meta}[h].\text{cnt}, c, c+1)$}
                \State Write payload $x$ to slab\_data[$h$][$c$]
                \State Write identifier to slab\_id\_buffer[$h$][$c$]
                \State $\mathcal{T}(v_{id}) \gets \langle h, c \rangle$
                \State \verb|__threadfence()|
                \State \texttt{atomicOr}(\&\text{slab\_meta}[h].\text{valid\_bitmap}, mask)
                \State \Return
            \EndIf
        \EndIf
    \EndIf
    \State $t \gets \texttt{atomicSub}(P_{top}, 1)$
    \If{$t \le 0$}
        \State \texttt{atomicAdd}$(P_{top}, 1)$
        \State \Return
    \EndIf
    \State $s_{new} \gets \text{free\_list}[t-1]$
    \State Set \text{slab\_meta}[$s_{new}$].\text{valid\_count} $\gets 1$
    \State Set \text{slab\_meta}[$s_{new}$].\text{validity\_bitmap} $\gets 0$
    \State Set \text{slab\_meta}[$s_{new}$].\text{next} $\gets h$
    \State \verb|__threadfence()|
    \If{$\texttt{atomicCAS}(\&H[\ell], h, s_{new}) == h$}
        \State Write payload $x$ to slab\_data[$s_{new}$][$0$]
        \State Write identifier to slab\_id\_buffer[$s_{new}$][$0$]
        \State $\mathcal{T}(v_{id}) \gets \langle s_{new}, 0 \rangle$
        \State \verb|__threadfence()|
        \State \texttt{atomicOr(\&slab\_meta[s\_new].validity\_bitmap)}
        \State \Return
    \Else
        \State $t_{ret} \gets \texttt{atomicAdd}(P_{top}, 1)$
        \State $\text{free\_list}[t_{ret}] \gets s_{new}$        
    \EndIf
\EndWhile
\end{algorithmic}
\end{algorithm}

\subsection{Coalesced Search on Non-Contiguous Slabs}
\label{sec:methods_search_kernel}

SIVF stores each inverted list as a linked chain of fixed-capacity slabs, which replaces contiguous scans with traversal via
\texttt{next}. To retain high GPU throughput, we use a warp-cooperative search kernel that matches slab capacity to the warp
width ($C=32$). The kernel launches one warp per query. The warp first stages the query vector into shared memory, then probes
$nprobe$ coarse lists returned by the quantizer. For each probed list $\ell$, the warp traverses the slab chain starting from
$s=H[\ell]$. At each slab $s$, lane $j$ consults \texttt{slab\_meta[$s$].validity\_bitmap} and evaluates slot $j$ only if the
corresponding bit is set. If set, the lane loads the payload from \texttt{slab data[$s$][$j$]}, retrieves the label from
\texttt{slab\_ids[$s$][$j$]}, computes the distance to the shared query, and updates a small per-lane top-$k$ structure kept in
registers. After traversal, lanes write their local top-$k$ candidates to shared memory and one lane merges them into the final
top-$k$ result for the query. To ensure termination under unexpected corruption, traversal is bounded and the kernel breaks on
self-loops. 

Algorithm~\ref{alg:sivf_search} summarizes the warp cooperative search procedure. To maximize throughput and memory coalescence, the search kernel assigns exactly one hardware warp (32 threads) to process each individual query. The warp begins by staging the query vector into shared memory, ensuring fast and repeated access for distance computations. Concurrently, each thread initializes a thread local top-$k$ priority queue stored entirely in fast registers.
The warp then iterates over the $nprobe$ candidate lists identified by the coarse quantizer. For each valid list, the threads cooperatively traverse the linked chain of fixed capacity slabs. At each slab, the workload is distributed perfectly across the warp: thread $j$ evaluates the $j$-th slot. The thread first inspects the corresponding bit in the validity bitmap. If the bit is set, indicating a valid vector, the thread loads the payload, computes the distance to the shared query vector, and updates its local top-$k$ queue. The traversal includes explicit guards against infinite loops by checking traversal bounds and self referential pointers.
Upon exhausting the candidate lists or reaching the traversal limit, the warp transitions to a reduction phase. All 32 threads write their local top-$k$ candidates into shared memory. Finally, a single designated thread merges these partial lists to formulate the definitive global top-$k$ results for the query, writing the final distances and labels to the output buffer.

\begin{algorithm}[t]
\caption{Warp-cooperative search in SIVF}
\label{alg:sivf_search}
\begin{algorithmic}[1]
\State Input: queries $Q$, probed list ids $coarse\_ids$, head array $H[\cdot]$
\State Input: slab metadata and data, buffer $slab\_ids$, $k$, $nprobe$
\State Output: top-$k$ distances and labels for each query

\For{each query index $q$ in parallel}
    \State Launch one warp for query $q$
    \State Copy $Q[q]$ into shared memory
    \State Initialize per-thread local top-$k$ lists in registers
    \For{$p = 0$ to $nprobe-1$}
        \State $\ell \gets coarse\_ids[q,p]$
        \If{$\ell$ is invalid} \State continue \EndIf
        \State $s \gets H[\ell]$
        \While{$s$ is valid and traversal bound not exceeded}
            \State $md \gets slab\_meta[s]$
            \If{$md.next = s$} \State break \EndIf
            \If{$j < 32$ and $md.validity\_bitmap[j]$ is set}
                \State Load vector $\mathbf{x}_{s,j}$ from slab data
                \State Compute $d(\mathbf{q}, \mathbf{x}_{s,j})$
                \State $id \gets slab\_ids[s,j]$
                \State Insert $(d,id)$ into local top-$k$
            \EndIf
            \State $s \gets md.next$
        \EndWhile
    \EndFor
    \State Write local top-$k$ lists to shared memory
    \State One thread merges 32 lists and writes final top-$k$ outputs
\EndFor
\end{algorithmic}
\end{algorithm}

\subsection{Lazy Eviction via Address Translation Table}
\label{sec:bitmap_lazy_eviction}

SIVF supports constant-time deletion by combining an Address Translation Table (ATT) with bitmap-based lazy eviction.
The ATT $\mathcal{T}$ maps each identifier $u$ to its physical coordinate in the slab pool as a 64-bit value
$\mathcal{T}[u] = (\text{idx}_{slab} \ll 32)\ \mid\ \text{idx}_{slot}$,
or \texttt{INVALID} if absent. Given $\mathcal{T}[u]$, a deletion thread decodes $(\text{idx}_{slab},\text{idx}_{slot})$ and
clears the corresponding bit in the slab validity bitmap using an atomic read-modify-write
(\texttt{atomicAnd} with $mask=\sim(1\ll \text{idx}_{slot})$). To handle duplicates and races, the kernel uses the pre-update bitmap value
to detect a $1\rightarrow 0$ transition. Bookkeeping and reclamation occur strictly on this transition: the thread decrements
the slab's \texttt{valid\_count} and marks $\mathcal{T}[u]$ as \texttt{INVALID}. If the decremented \texttt{valid\_count}
drops to zero, implying the slab has become fully empty, the thread atomically pushes the slab index back to the global free stack
$P_{top}$. This immediate reclamation strategy allows SIVF to reuse memory for new insertions without requiring heavy background compaction.

Algorithm~\ref{alg:sivf_delete} summarizes the lazy eviction procedure with slab level reclamation. The deletion process executes in parallel across a batch of target identifiers. For each identifier, the thread performs a constant time lookup in the Address Translation Table to retrieve the physical coordinate. If the coordinate is valid, the thread decodes the 64-bit integer into a slab index and a slot index.
Logical deletion is achieved by generating a bitmask and applying an atomic AND operation to clear the corresponding bit in the slab validity bitmap. The thread captures the previous state of the bitmap to verify if the bit was originally set. This verification ensures idempotence and prevents redundant bookkeeping when multiple threads attempt to delete the same vector. If the thread successfully transitions the bit from 1 to 0, it proceeds to atomically decrement the occupancy counter of the slab and invalidates the entry in the Address Translation Table. If the occupancy counter drops to zero, the thread instantly reclaims the empty slab by atomically pushing its index back onto the global free list and resetting its bitmap, making the memory immediately available for future insertions.

\begin{algorithm}[t]
\caption{Lazy eviction with slab-wise reclamation in SIVF}
\label{alg:sivf_delete}
\begin{algorithmic}[1]
\State Input: deletion identifiers $U=\{u_1,\ldots,u_m\}$
\State Global: Address Table $\mathcal{T}$, slab metadata, free list, stack pointer $P_{top}$

\For{each $u$ in $U$ in parallel}
    \State $coord \gets \mathcal{T}[u]$
    \If{$coord == \text{INVALID}$} \State \textbf{continue} \EndIf
    \State $\text{idx}_{slab} \gets coord \gg 32$
    \State $\text{idx}_{slot} \gets coord \mathbin{\&} (2^{32}-1)$
    \State $mask \gets \quad \sim(1 \ll \text{idx}_{slot})$
    \State $old\_map \gets \texttt{atomicAnd}(\text{slab}[\text{idx}_{slab}].\text{bitmap}, mask)$
    \If{$((old\_map \gg \text{idx}_{slot}) \mathbin{\&} 1) == 1$}
        \State $old\_cnt = \texttt{atomicSub}(\&\text{slab\_meta}[\text{idx}_{slab}].\text{cnt}, 1)$
        \State $\mathcal{T}[u] \gets \text{INVALID}$
        \If{$old\_cnt == 1$}
             \State $t \gets \texttt{atomicAdd}(P_{top}, 1)$
             \State $\text{free\_list}[t] \gets \text{idx}_{slab}$
             \State $\text{slab\_meta}[\text{idx}_{slab}].\text{bitmap} \gets 0$
        \EndIf
    \EndIf
\EndFor
\end{algorithmic}
\end{algorithm}

\subsection{Theoretical Analysis}
\label{sec:theoretical_analysis}

\subsubsection{Correctness}
\label{sec:correctness}

We prove three properties. The first establishes the correctness of parallel ingestion in Algorithm~\ref{alg:sivf_ingestion}.
The second shows that search in Algorithm~\ref{alg:sivf_search} is safe under concurrent ingestion and deletion.
The third proves that lazy eviction in Algorithm~\ref{alg:sivf_delete} is linearizable with a clear linearization point.

We use the same state as the pseudocode. For each list id $\ell$, $H[\ell]$ stores the head slab id. Each slab $s$ has
metadata \texttt{slab\_meta[$s$]} including \texttt{next}, \texttt{valid\_count}, and a 32-bit
\texttt{validity\_bitmap}. A slot $(s,o)$ is logically present if and only if bit $o$ in the bitmap is 1.
Payloads and ids are stored in \texttt{slab\_data[$s$][$o$]} and \texttt{slab\_ids[$s$][$o$]}.
The address table $\mathcal{T}$ maps a vector id $u$ to a coordinate $(s,o)$ encoded in \texttt{coord}, or \texttt{INVALID}.

\begin{theorem}[Parallel ingestion is safe and linearizable]
Consider any concurrent execution of Algorithm~\ref{alg:sivf_ingestion}.
For every insertion that returns successfully, there exists a unique slot $(s,o)$ such that:
(i) the insertion operation writes the payload and id into that slot, then sets bit $o$ in \textup{\texttt{validity\_bitmap}} exactly once,
(ii) once the bit is set, the slot contains a fully initialized payload and id, and
(iii) the insertion can be linearized at the atomic bit set operation \textup{\texttt{atomicOr}} that makes the slot visible.
\end{theorem}

\begin{proof}
See Appendix, Theorem~\ref{thm:ingestion}.
\end{proof}

\begin{theorem}[Search safety under concurrent ingestion and deletion]
In any concurrent execution of Algorithms~\ref{alg:sivf_ingestion}, \ref{alg:sivf_search}, and \ref{alg:sivf_delete},
every payload and id read performed by Algorithm~\ref{alg:sivf_search} is fully initialized.
\end{theorem}

\begin{proof}
See Appendix, Theorem~\ref{thm:search}.
\end{proof}

\begin{theorem}[Lazy eviction is linearizable and makes deleted vectors invisible]
Consider any concurrent execution of Algorithms~\ref{alg:sivf_search} and \ref{alg:sivf_delete}.
For any delete request on $u$ such that 
$\mathcal{T}[u] \neq \textnormal{\texttt{INVALID}}$
and decodes to
$(\text{idx}_{slab},\text{idx}_{slot})$, define
\[
mask = \mathbin{\sim}(1 \ll \text{idx}_{slot}).
\]
Then the atomic operation
$old\_map \gets atomicAnd()$ 
in Algorithm~\ref{alg:sivf_delete} constitutes the linearization point for logical deletion. After this atomic operation takes effect,
the slot becomes logically absent from the search space. Repeated deletes of the same id are idempotent and safe.
\end{theorem}

\begin{proof}
See Appendix, Theorem~\ref{thm:delete}.
\end{proof}

\subsubsection{Time Complexity}
We analyze the time complexity relative to the total number of vectors $N$, the number of lists $n_{list}$, and dimensionality $D$. For ingestion, SIVF guarantees strict $O(1)$ time complexity as it requires only atomic counter updates and pointer manipulation to append data, independent of the current list size. Similarly, deletion achieves $O(1)$ complexity by utilizing the Address Translation Table for constant-time lookup followed by a single atomic bitmap invalidation, completely eliminating the need for memory compaction or data movement. For search, the complexity remains $O(nprobe \cdot (N/n_{list}) \cdot D)$, where $nprobe$ is the user-configured number of lists to scan; while the linked-slab layout introduces minor pointer-chasing overhead, it preserves the asymptotic complexity class of inverted file search.

\subsubsection{Space Complexity}
The total space complexity is dominated by the vector payload, scaling linearly as $O(N \cdot D)$. The structural overhead consists of the Address Translation Table ($O(N)$, 8 bytes per entry) and slab metadata. Since metadata is amortized over the fixed slab capacity ($C=32$), its footprint is negligible compared to high-dimensional payloads. Regarding memory efficiency, unlike dynamic arrays that typically reserve up to $2 \times$ capacity to amortize resizing costs, SIVF grows in fine-grained increments of single slabs. Although lazy deletion introduces sparse internal fragmentation (unused slots within active slabs), the immediate reclamation of fully empty slabs bounds this waste, ensuring the total memory footprint remains proportional to the active dataset size.

\section{System Implementation}
\label{sec:implementation_details}

We implement SIVF as a fully integrated extension to the GPU components of the widely-adopted Faiss library~\cite{douze2024faiss}. Our implementation represents a significant systems engineering effort, involving approximately 13,000 lines of code as of January 2026. The majority of this codebase consists of optimized C++ and CUDA system implementation (spanning the core Faiss integration, slab memory management, and distributed MPI runtime), while the remaining lines comprise Python and Bash utility scripts for experimental orchestration. We have open-sourced the complete implementation and evaluation scripts on GitHub.

\subsection{Core Classes}

The core architecture centers on the \texttt{GpuIndexSIVF} class, a direct subclass of Faiss's \texttt{GpuIndex}. This class integrates our custom \texttt{SlabManager} to orchestrate dynamic memory on the GPU. Unlike standard static models, \texttt{SlabManager} maintains a monolithic memory pool and a device-side free list stack, enabling both warp-aligned allocation and $O(1)$ slab recycling without host intervention. To support high-concurrency operations, we implemented specialized GPU-native kernels for insertion, deletion, and search (encapsulated in \texttt{SIVFAppend}, \texttt{SIVFDelete}, and \texttt{SIVFSearch}). These operators interact with lock-free metadata structures via atomic state transitions to ensure data consistency during simultaneous read and write access.

Integrating these dynamic components into the static resource management of Faiss presented multiple challenges. Unlike standard Faiss indices which rely on flat arrays managed through the \texttt{DeviceTensor} class, SIVF implements its own allocator to minimize fragmentation overhead. We reconcile this by overriding the virtual \texttt{addImpl} and \texttt{searchImpl} methods, allowing SIVF to serve as a drop-in replacement in existing pipelines. Furthermore, our design adheres to the \texttt{GpuResources} context provided by Faiss. 

\subsection{Distributed Scale-out Architecture}
\label{sec:distributed_arch}

To support large-scale datasets exceeding single-device memory capacity, we extended SIVF with a shared-nothing, data-parallel distributed architecture. 

The global dataset is partitioned into disjoint shards, distributed across $P$ GPU workers. Unlike model-parallel approaches that split inverted lists across devices (which incurs heavy synchronization overheads during ingestion), SIVF employs \textit{Data Sharding}. Incoming vectors are routed to workers via a deterministic partitioning capabilities (e.g., round-robin or hash-based routing). This allows each GPU to ingest data into its local SIVF index independently, enabling the aggregate ingestion throughput to scale linearly with the cluster size, as evidenced in our evaluation (Section~\ref{sec:eval_multigpu}).

Query processing follows a \textit{Scatter-Gather} pattern. The client broadcasts the query vector to all workers via MPI. Each worker performs a warp-cooperative search on its local shard to retrieve the top-$k$ candidates. A global reduction step (implemented via \texttt{MPI\_Gather} or tree-based reduction) merges the partial results to produce the final global top-$k$. 
For deletion, since SIVF maintains a dense ID space, deletion requests are broadcast to all workers. Each worker checks its local Address Translation Table (ATT). Due to the disjoint partitioning, the target ID exists on at most one worker, which then executes the lock-free lazy eviction locally.

\section{Evaluation}
\label{sec:evaluation}



\subsection{Experimental Setup}
\label{sec:exp_setup}

To ensure a comprehensive evaluation, we adopt a two-tiered baseline strategy. 
First, our primary baseline is the standard Faiss GPU IVF, which serves as a direct architectural counterpart; this allows us to isolate the performance gains specifically attributable to our proposed memory management innovations, excluding confounding factors from differing algorithmic paradigms. 
Second, to position SIVF within the broader indexing landscape, we in Section~\ref{sec:eval_non_ivf} compare against representative non-IVF indices, including GPU CAGRA~\cite{10597683} (dynamic graph), GPU Flat (brute-force), HNSW~\cite{malkov2018efficient} (dynamic graph), NSG~\cite{10.14778/3303753.3303754} (static graph), and LSH~\cite{10.1145/997817.997857} (hashing).

Most experiments were conducted on a bare-metal node from the Chameleon Cloud testbed~\cite{keahey2020lessons} (CHI@UC site). The platform is equipped with dual-socket Intel Xeon Gold 6126 CPUs (Skylake microarchitecture), providing a total of 24 physical cores (48 threads with Hyper-Threading) clocked at 2.60 GHz. The host memory consists of 192 GiB of RAM. 
For acceleration, the node features an NVIDIA Quadro RTX 6000 GPU with 24 GB of GDDR6 VRAM. The system runs on a Ubuntu 24.04 environment with the CUDA toolkit installed: Driver ver. 560.35.05 and CUDA ver. 12.6.
Each reported data point represents the average of at least three independent executions. Error bars are omitted in figures where the standard deviation is negligible to maintain visual clarity. 

\subsection{Microbenchmarks}

\subsubsection{Ingestion}
\label{sec:eval_add_performance}

\begin{figure*}[t]
    \centering
    \includegraphics[width=\linewidth]{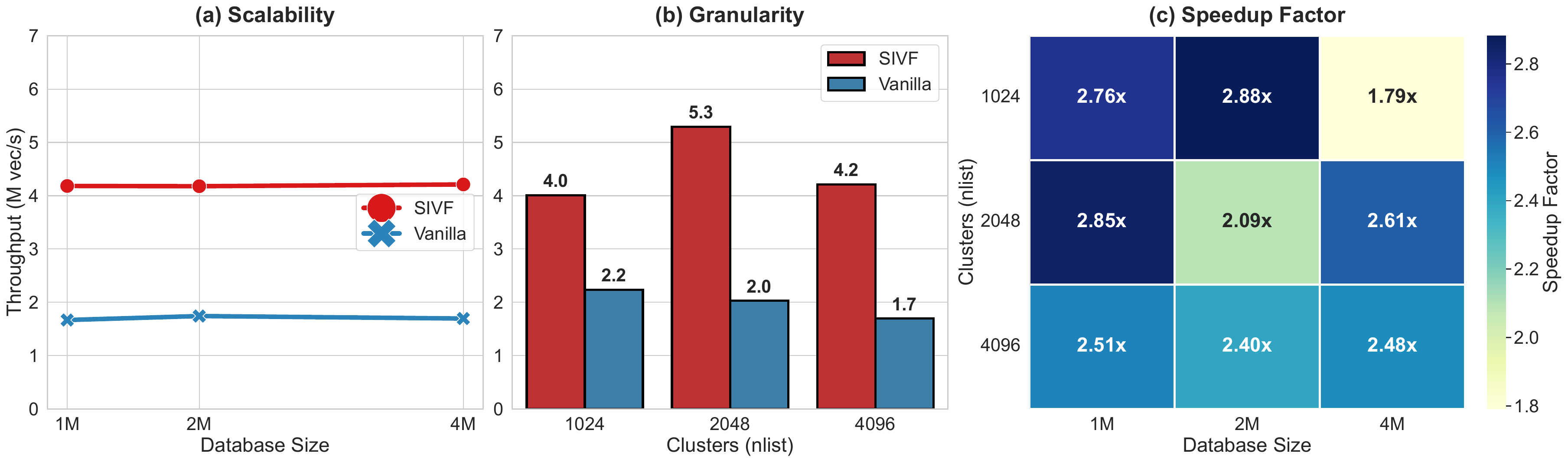}
    \caption{Microbenchmark performance of vector ingestion.}
    \label{fig:add_perf}
    \Description{Three performance graphs comparing SIVF and Vanilla Faiss.}
\end{figure*}

We evaluate ingestion throughput against the standard Faiss GPU IVFFlat baseline, varying database size ($N_B$: 1M--4M) and cluster count ($n_{list}$: 1024--16384). As shown in Figure~\ref{fig:add_perf}, SIVF achieves consistent performance advantages with up to 2.65$\times$ speedup.

As shown in Fig.~\ref{fig:add_perf}a, SIVF maintains a stable throughput of $\approx$4.2M vec/s as $N_B$ increases, significantly outperforming the baseline ($\approx$1.7M vec/s). This stability validates our lock-free SlabManager, which ensures $O(1)$ insertion cost independent of index occupancy, thereby avoiding the contiguous memory resizing overheads inherent to the baseline.
While the baseline throughput degrades monotonically as $n_{list}$ increases (dropping from 2.2M to 1.7M vec/s), Fig.~\ref{fig:add_perf}b shows that SIVF exhibits a performance sweet spot at $n_{list}=2048$, peaking at 5.3M vec/s. At this optimal granularity, SIVF achieves a $2.65\times$ speedup over the baseline. Even at high $n_{list}=4096$, where scattered memory writes typically hinder performance, SIVF maintains robust throughput at 4.2M vec/s, sustaining a $2.47\times$ advantage over the baseline's 1.7M vec/s.
As shown in Fig.~\ref{fig:add_perf}c, SIVF outperforms the baseline across all configurations, with speedups typically ranging from 2.4$\times$ to 2.9$\times$. Even under maximum contention (4M vectors, $n_{list}=1024$), SIVF retains a 1.79$\times$ advantage, confirming the efficiency of the non-blocking, slab-based pipeline for streaming scenarios.

\subsubsection{Deletion}
\label{sec:eval_deletion_performance}

We evaluate the efficiency of the vector deletion mechanism in SIVF by comparing it against the standard Faiss GPU IVF baseline. The experiment measures latency and throughput when removing a batch of 10,000 vectors from a populated index of one million 128-dimensional vectors.

\begin{figure}[t]
    \centering
    \includegraphics[width=\columnwidth]{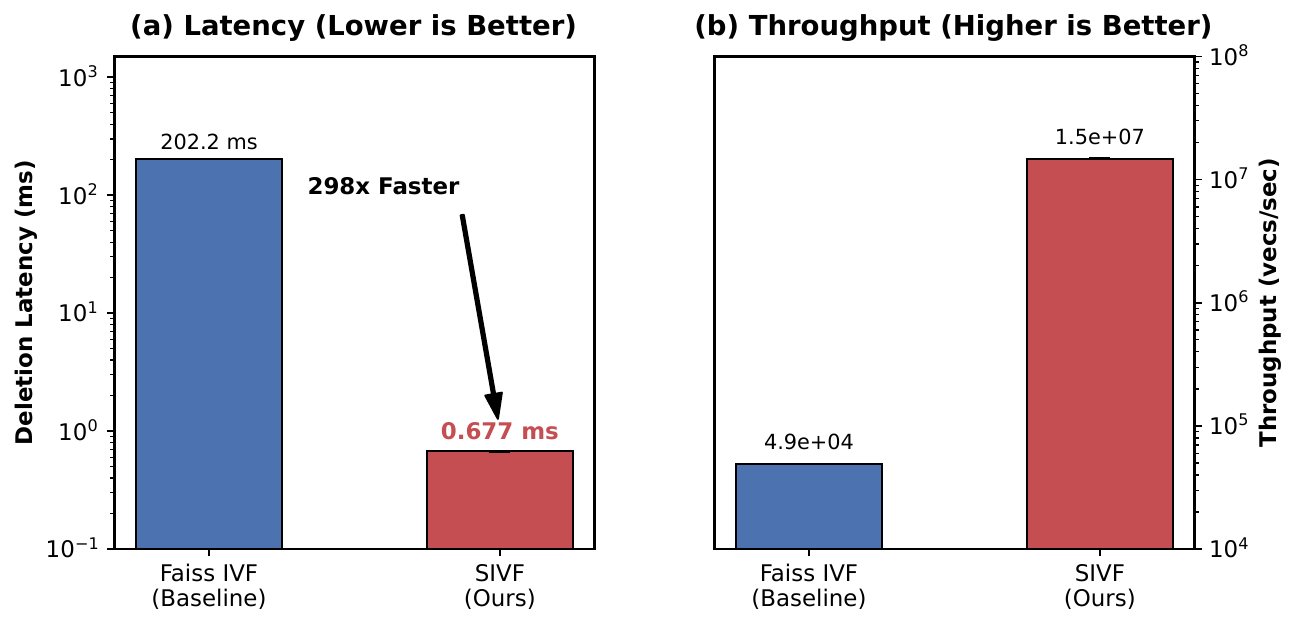}
    \caption{Microbenchmark performance of vector deletion. 
    }
    \label{fig:deletion_perf}
    \Description{Two bar graphs comparing deletion latency and throughput between Faiss IVF and SIVF, highlighting the significant speedup achieved by SIVF.}
\end{figure}

Figure~\ref{fig:deletion_perf} illustrates the performance comparison between the two methods. The results indicate that the baseline Faiss implementation incurs a substantial deletion latency of approximately 202.2 ms. In stark contrast, SIVF completes the same batch deletion operation in an average of 0.68 ms. This represents a massive speedup of 298.5$\times$. Correspondingly, deletion throughput surges from $4.9 \times 10^4$ vec/s in the baseline to nearly $1.5 \times 10^7$ vec/s in SIVF.


\subsubsection{Parameter Sensitivity and Ablation Study}
\label{sec:eval_param_sensitivity}

\begin{figure*}[t]
    \centering
    \begin{minipage}{0.48\textwidth}
        \centering
        \includegraphics[width=\linewidth]{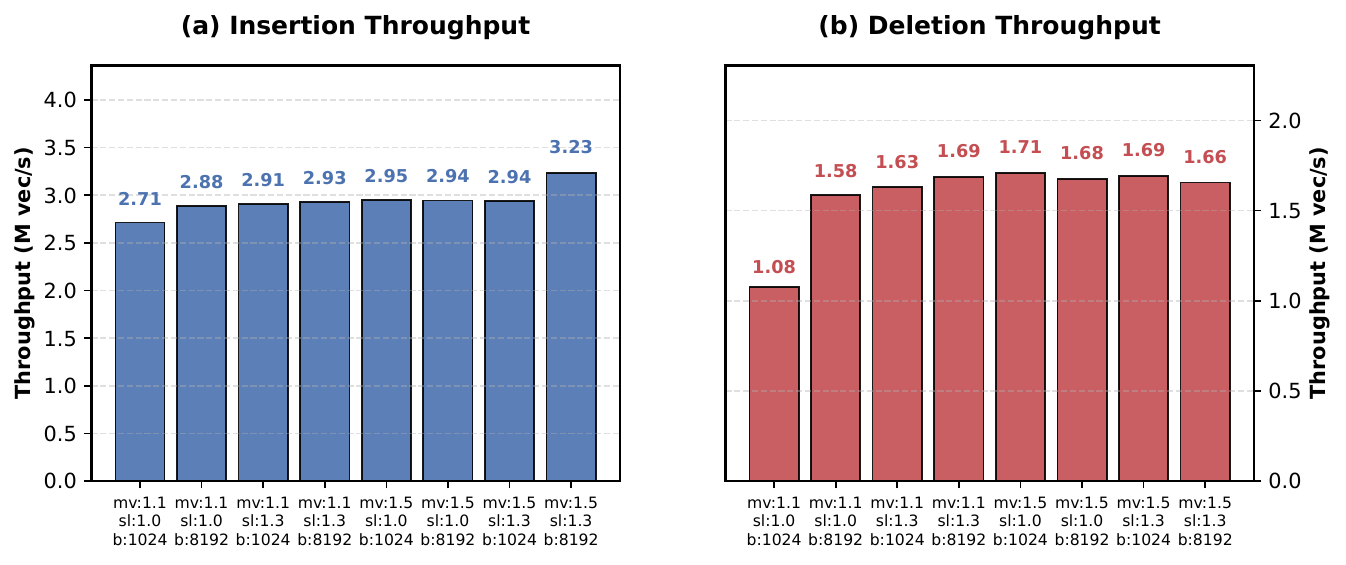}
        \caption{Throughput sensitivity analysis.}
        \Description{Two performance graphs showing throughput sensitivity analysis for SIVF.}
        \label{fig:sivf_sensitivity_qps}
    \end{minipage}
    \hfill
    \begin{minipage}{0.48\textwidth}
        \centering
        \includegraphics[width=\linewidth]{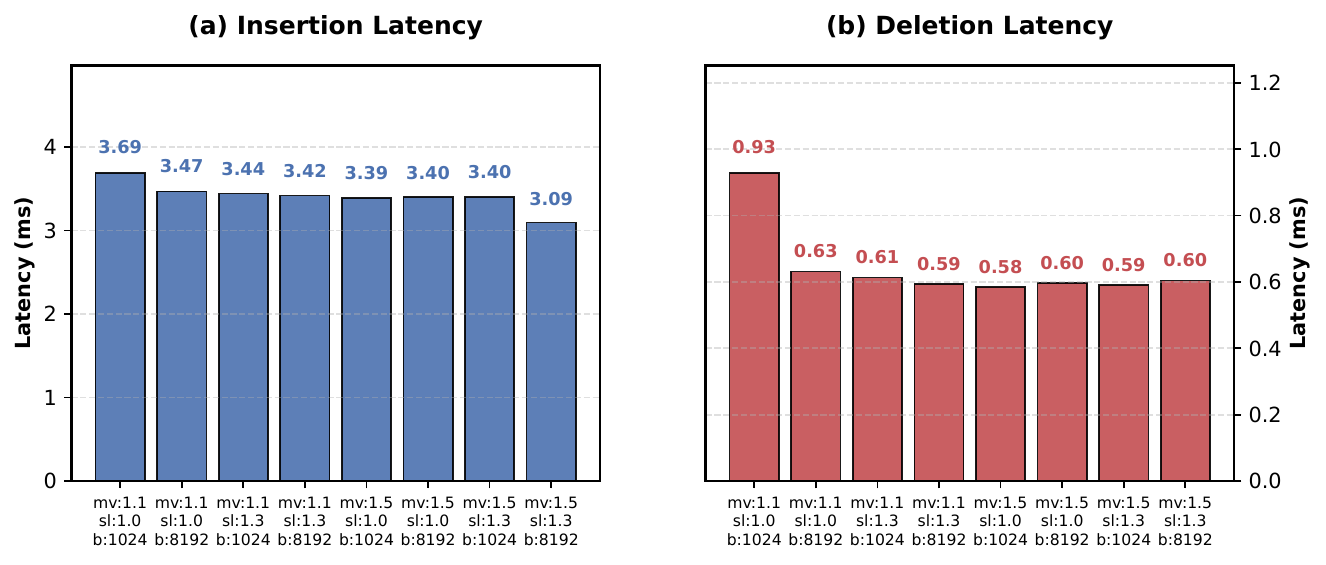}
        \caption{Latency sensitivity analysis.}
        \Description{Two performance graphs showing throughput and latency sensitivity analysis for SIVF.}
        \label{fig:sivf_sensitivity_lat}
    \end{minipage}
\end{figure*}

We evaluate system robustness by sweeping three parameters: (i) vector capacity factor ($maxvec\_factor$, or $mv$), (ii) slab pool redundancy ($slab\_factor$, or $sl$), and (iii) deletion batch size ($b$). These control logical headroom, pre-allocated memory for bursts, and the trade-off between amortization and freshness. Figures~\ref{fig:sivf_sensitivity_qps} and \ref{fig:sivf_sensitivity_lat} summarize the results.

Figures~\ref{fig:sivf_sensitivity_qps} and \ref{fig:sivf_sensitivity_lat} illustrate the impact of pre-allocation strategies and batch sizes on system performance. Insertion throughput correlates positively with memory provisioning, peaking at 3.23M vec/s with a $maxvec\_factor$ of 1.5 and larger batches, as generous pre-allocation minimizes dynamic resizing and contention. While deletion throughput is sensitive to batching under tight constraints (dropping to 1.08M vec/s), increasing the $slab\_factor$ or $maxvec\_factor$ effectively decouples performance from resource limitations, stabilizing throughput at $\approx$1.7M vec/s. End-to-end latency remains robust across all settings, with insertions ranging from 3.09--3.69 ms and deletions achieving sub-millisecond performance (0.58--0.93 ms). This $3\times$--$5\times$ speed advantage for deletions stems from SIVF's lock-free bitmap mechanism, while the observed latency reduction with larger batches confirms that amortizing kernel launch overheads is a key driver for high-velocity streaming mutations.

\subsection{Real-world Datasets}
\label{sec:eval_real_world}

We benchmark SIVF using four popular real-world datasets representing diverse modalities and varying degrees of cluster skewness: Deep1B~\cite{babenko2016efficient} (96 dimensions, deep features, imbalance factor $\mathcal{I}=1.23$), SIFT1M~\cite{jegou2011product} (128 dimensions, vision, $\mathcal{I}=1.24$), T2I-1B~\cite{simhadri2025results} (200 dimensions, multimodal/RAG, $\mathcal{I}=1.21$), and GIST1M~\cite{jegou2011product} (960 dimensions, high-dim features, $\mathcal{I}=1.76$). The high imbalance factor of GIST1M, in particular, stresses the allocator's ability to handle non-uniform distributions. While the subset size (1M vectors) fits within GPU memory, these experiments validate performance within the \textit{active window} of streaming applications. Section~\ref{sec:eval_streaming} will report their stability under high-churn workloads.

\begin{figure*}[t]
    \centering
    \begin{minipage}{0.32\textwidth}
        \centering
        \includegraphics[width=\linewidth]{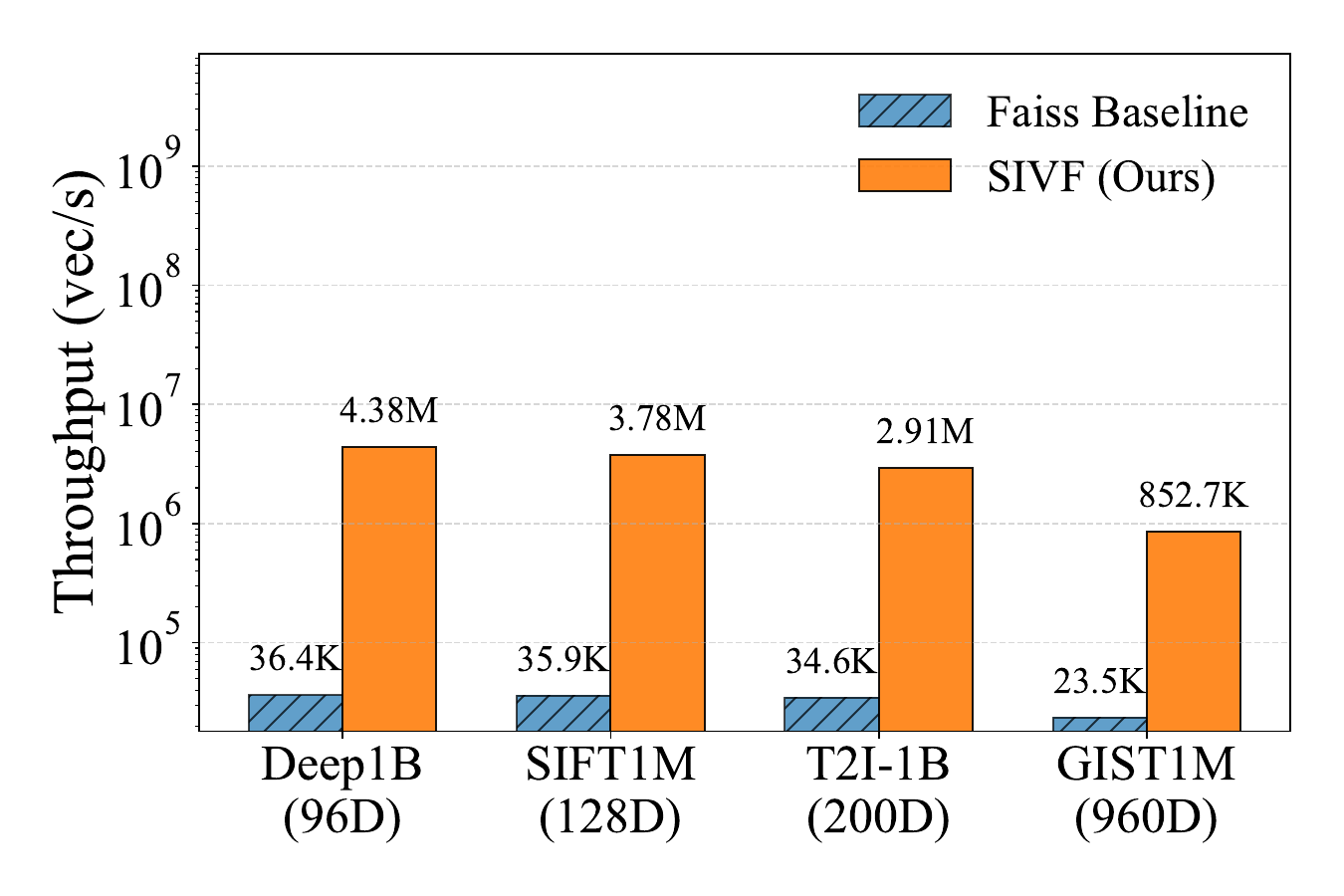}
        \caption{Ingestion throughput.}
        \Description{Bar graph comparing ingestion throughput between SIVF and Vanilla Faiss.}
        \label{fig:eval_ingestion}
    \end{minipage}
    \hfill 
    \begin{minipage}{0.32\textwidth}
        \centering
        \includegraphics[width=\linewidth]{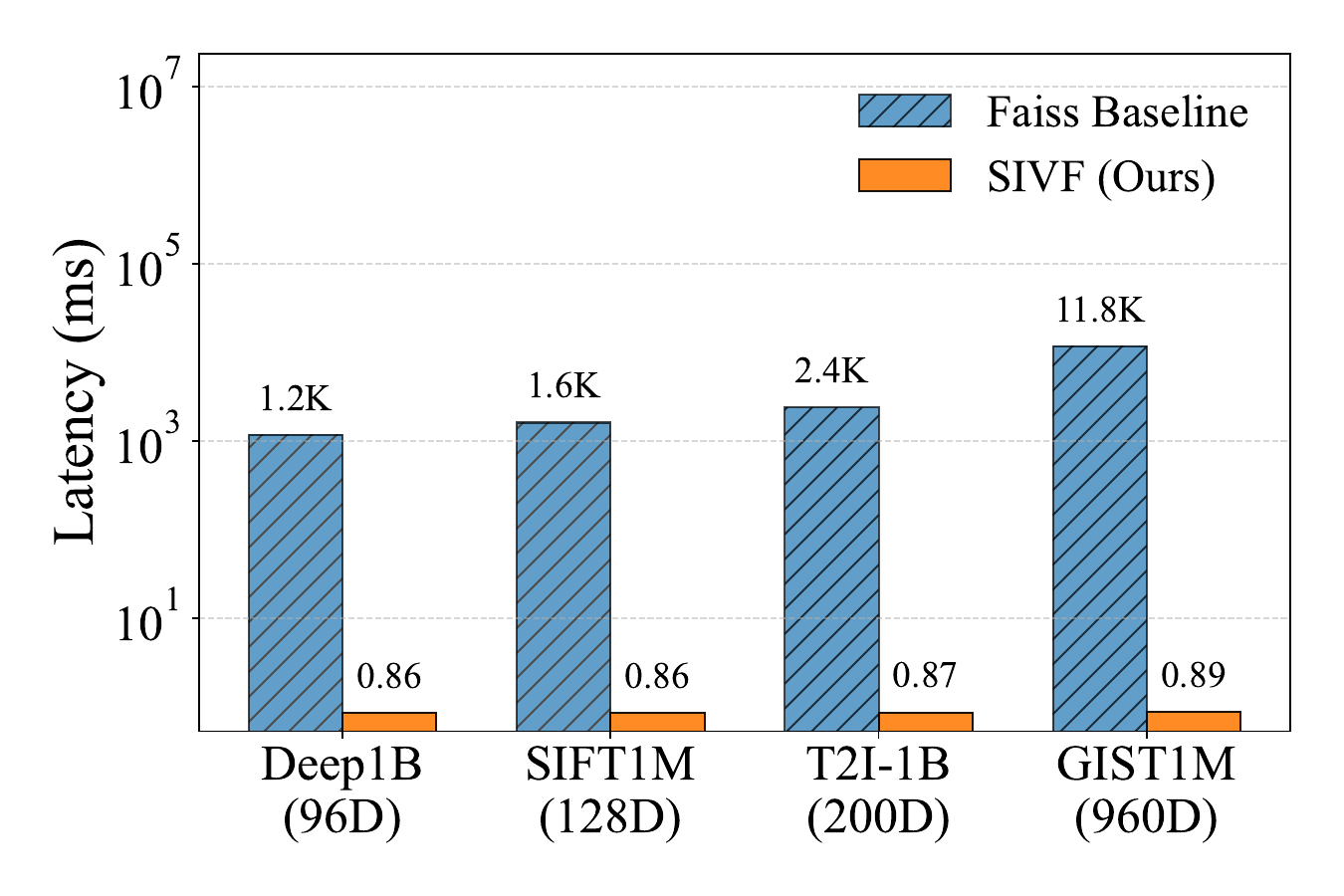}
        \caption{Deletion latency.}
        \Description{Bar graph comparing deletion latency between SIVF and Vanilla Faiss.}
        \label{fig:eval_deletion}
    \end{minipage}
    \hfill 
    \begin{minipage}{0.32\textwidth}
        \centering
        \includegraphics[width=\linewidth]{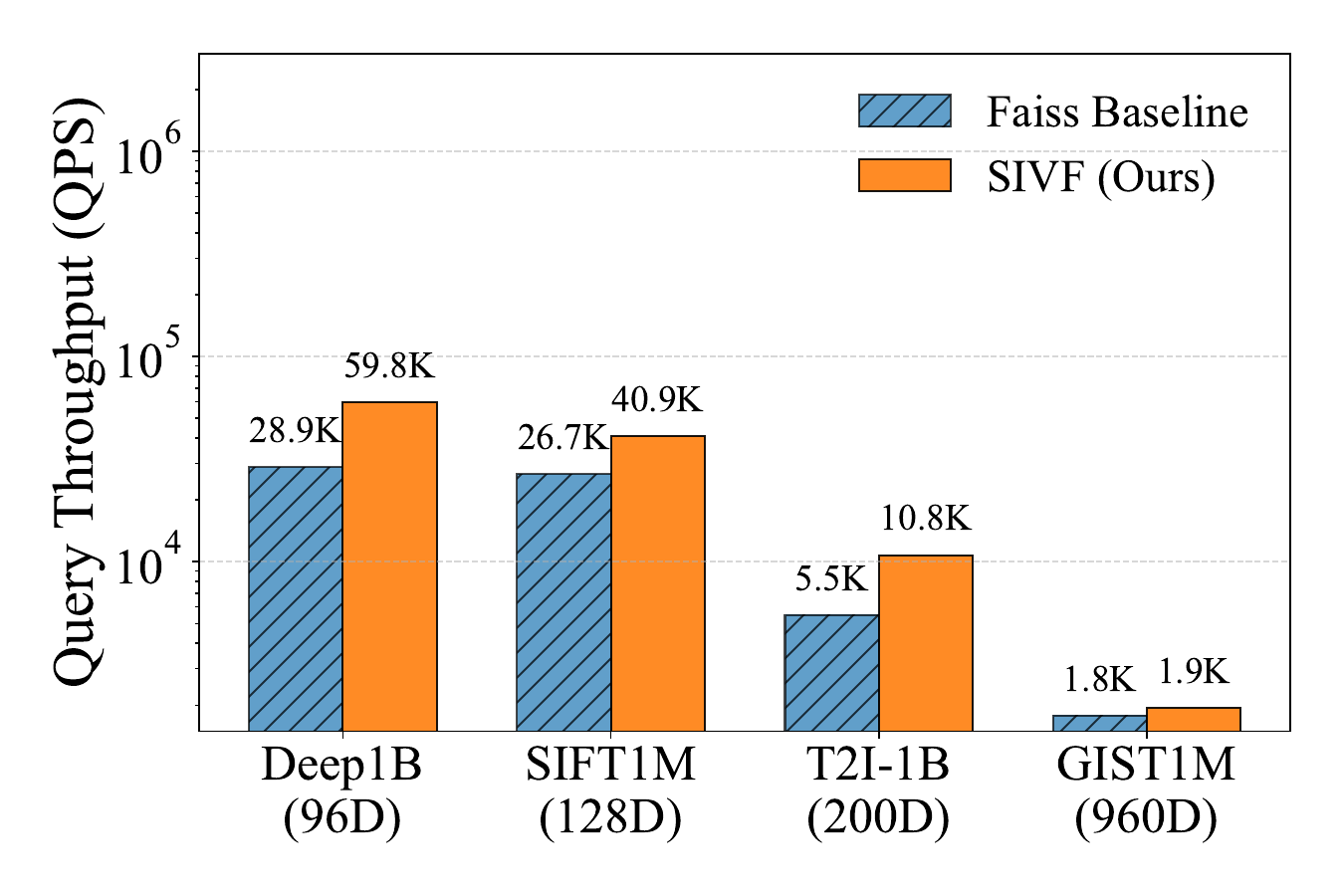}
        \caption{Search performance (QPS).}
        \Description{Bar graph comparing search performance (QPS) between SIVF and Vanilla Faiss.}
        \label{fig:eval_search}
    \end{minipage}
\end{figure*}

As shown in Figure~\ref{fig:eval_ingestion}, SIVF dominates ingestion scalability across all dimensions. On Deep1B, SIVF achieves a peak throughput of 4.38M vec/s ($120\times$ speedup over the baseline's 36.4K vec/s). This advantage persists on SIFT1M (3.78M vec/s, $105\times$) and T2I-1B (2.91M vec/s, $84\times$). Even on high-dimensional GIST1M, SIVF maintains 852.7K vec/s ($36\times$ speedup), confirming that our pre-allocated slab architecture effectively saturates GPU bandwidth by eliminating dynamic resizing overhead.

Figure~\ref{fig:eval_deletion} highlights the critical performance divergence. Baseline latency scales poorly due to CPU-GPU roundtrips, varying from 1.2s (Deep1B) to 11.8s (GIST1M). In contrast, SIVF's in-place bitmap updates decouple latency from dimension, maintaining sub-millisecond performance ($<0.9$ ms) across all datasets. Notably on GIST1M, SIVF reduces latency from 11,843 ms to 0.89 ms, a $13,307\times$ improvement, validating its feasibility for real-time streams.

Figure~\ref{fig:eval_search} compares query throughput across diverse datasets. 
The results demonstrate that SIVF fundamentally overcomes the performance bottleneck typical of dynamic indexing without compromising retrieval accuracy. 
On Deep1B, SIVF outperforms the baseline by 2.07x (59.8K vs 28.9K QPS). 
On SIFT1M, it retains a 1.5x lead (40.9K vs 26.7K QPS), while on T2I-1B, it provides competitive performance reaching 10.8K QPS. 
While microbenchmarks in sparse scenarios (e.g., $n_{list}=16384$) may show efficiency ratios between 0.40x and 0.52x due to the physical trade-off of pointer chasing, SIVF demonstrates that query performance on real world datasets is highly adjustable through the $n_{list}$ and $n_{probe}$ parameters. 
By optimizing the coarse quantizer granularity, such as setting $n_{list}=8192$ for GIST1M and $n_{list}=4096$ for T2I-1B, SIVF maintains parity or superior query throughput even as search reaches high recall regimes. 

\begin{figure}[t]
    \centering
    \includegraphics[width=\linewidth]{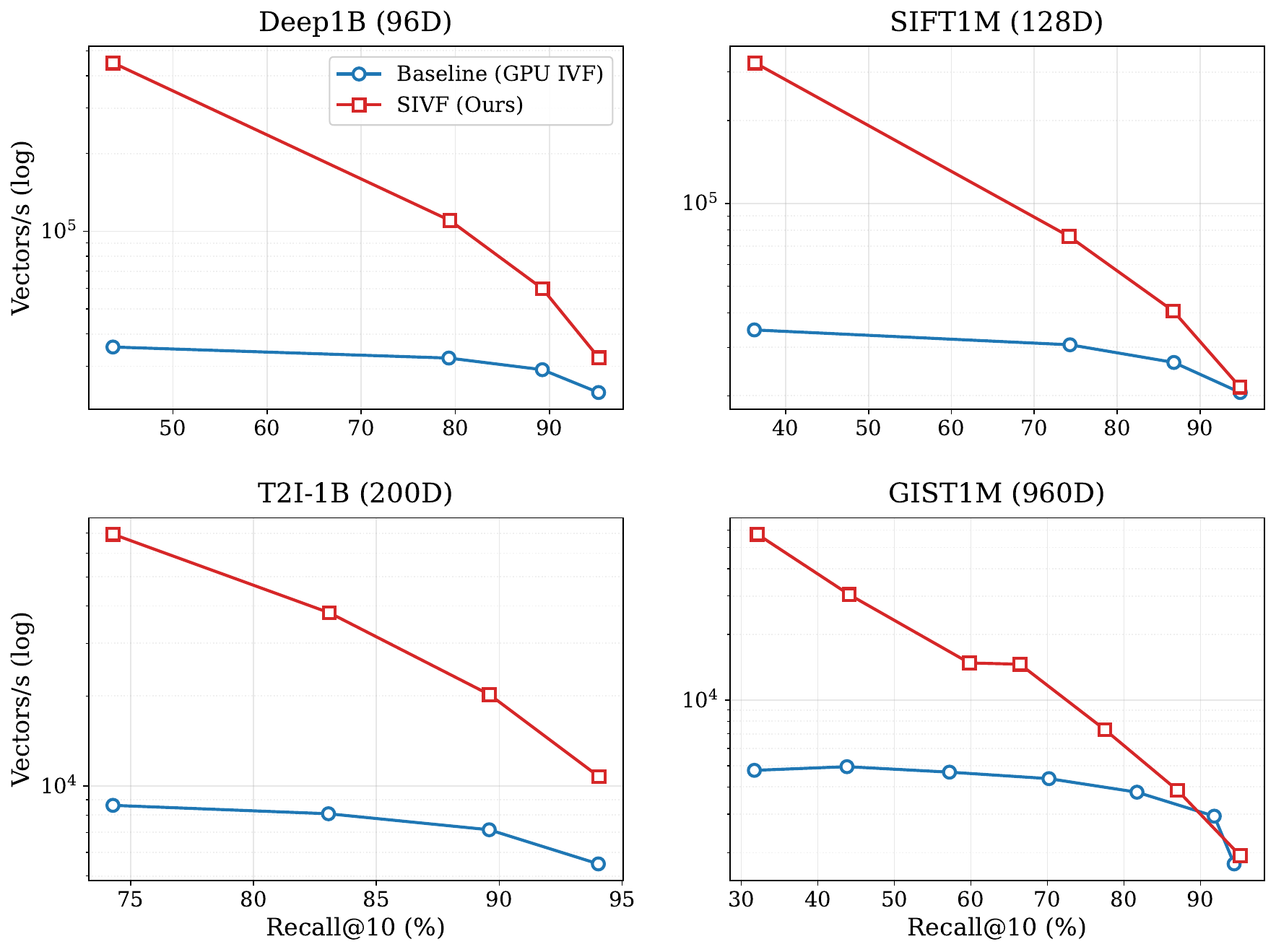}
    \caption{Throughput-Recall Pareto frontier analysis across four datasets (Deep1B, SIFT1M, T2I-1B, and GIST1M).}
    \Description{Four line graphs comparing QPS vs Recall@10 for SIVF and Vanilla Faiss across four datasets.}
    \label{fig:pareto}
\end{figure}

Figure~\ref{fig:pareto} illustrates the trade-off between query throughput and retrieval quality measured by Recall@10. 
The results confirm that SIVF provides a strictly superior operational envelope compared to the contiguous baseline across all tested dimensions. 
First, SIVF achieves strict recall parity, reaching identical maximum recall targets such as 95.2\% on GIST1M and 94.1\% on T2I-1B. 
This validates that the non-contiguous slab architecture introduces no precision loss. 
Second, SIVF maintains its throughput advantage even as the search depth increases to target high recall regimes. 
The slab management layer ensures high memory coalescing efficiency during vector traversal, which effectively masks the latency inherent in pointer chasing. 
Third, the consistent gap between the SIVF and baseline curves across the entire recall spectrum suggests that the system successfully eliminates the traditional trade-off between index flexibility and search performance. 

\subsection{Non-Uniform Data Distribution}
\label{sec:eval_nonuniform}

To evaluate system robustness under skewed workloads, we benchmark SIVF against Faiss IVFFlat and an advanced baseline, FluxVec, using a Zipfian distribution of one million vectors. FluxVec attempts to optimize ingestion by pre-sorting vectors assigned to each cluster for batched insertion. However, empirical results demonstrate that this sorting overhead outweighs the batching benefits.

\begin{figure}[t]
\centering
\includegraphics[width=1.0\linewidth]{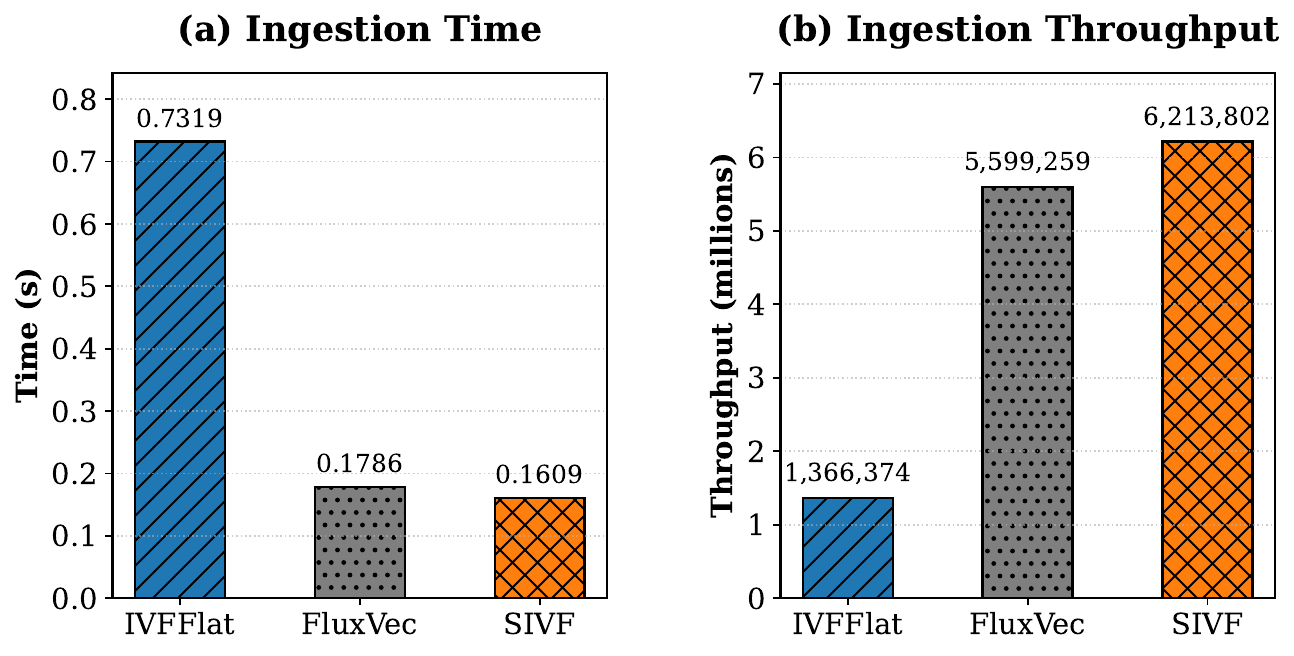}
\caption{Performance comparison under Zipfian skewed data distribution.}
\Description{Two bar graphs comparing ingestion time and query throughput among IVFFlat, FluxVec, and SIVF under skewed distribution.}
\label{fig:eval_nonuniform}
\end{figure}

As illustrated in Figure~\ref{fig:eval_nonuniform}, SIVF natively outperforms both methods without requiring explicit pre-sorting. SIVF completes the ingestion process in 0.1609 seconds, which is significantly faster than the 0.7319 seconds required by Faiss IVFFlat and the 0.1786 seconds needed by FluxVec. Furthermore, SIVF achieves a peak throughput of over 6.21 million QPS, outperforming FluxVec (5.59 million QPS) and Faiss (1.36 million QPS). This confirms that SIVF maintains high efficiency and stability even when the data distribution is highly skewed.

\subsection{End-to-End Streaming Performance}
\label{sec:eval_streaming}

\subsubsection{Sliding window}

We evaluate real-time applicability using a \textit{Sliding Window} benchmark that mimics production streams. The system maintains a fixed active window ($W$) by ingesting a new batch ($B$) and evicting the oldest batch ($B$) at each step. We test on SIFT1M ($W=200\text{K}, B=10\text{K}$) and GIST1M ($W=100\text{K}, B=5\text{K}$).

\begin{figure}[t]
    \centering
    \includegraphics[width=1.0\linewidth]{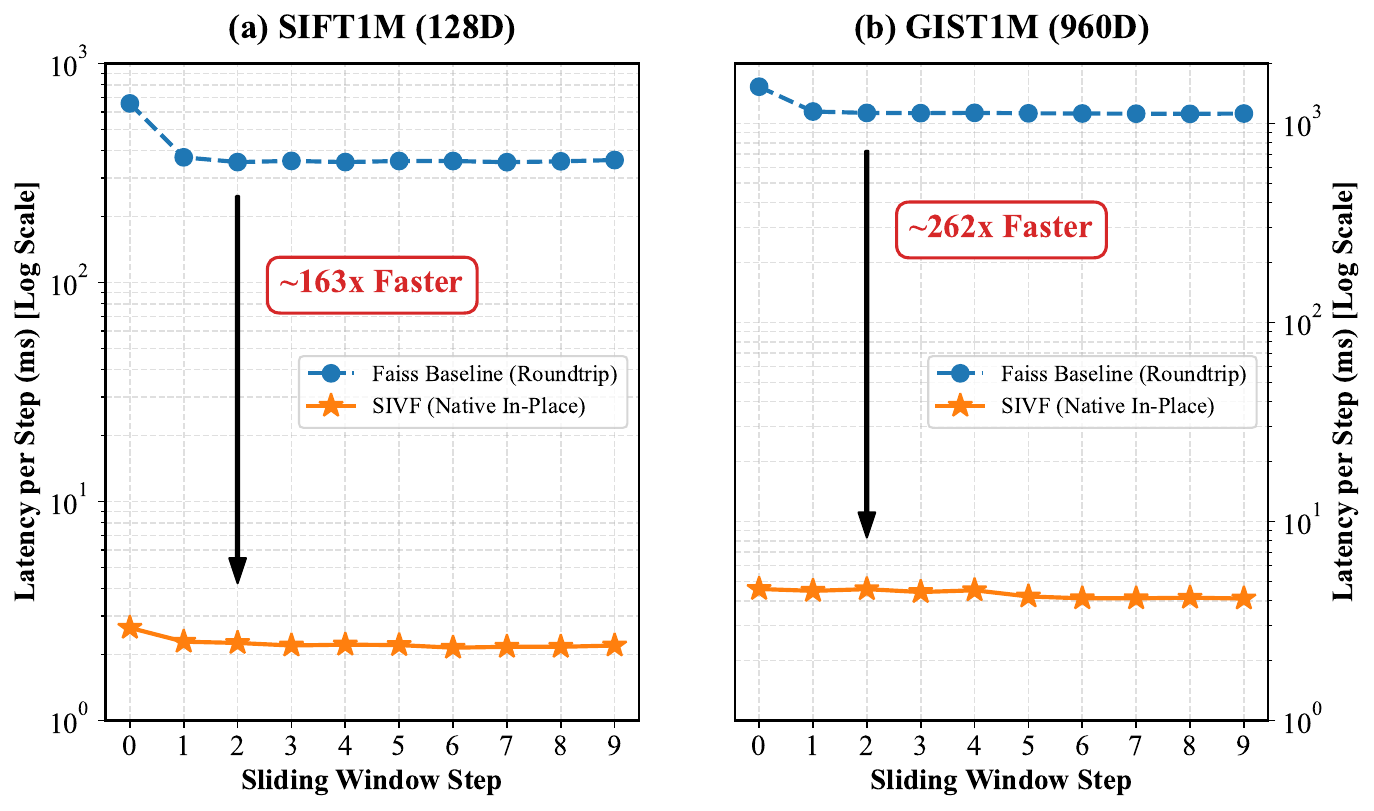}
    \caption{End-to-End Streaming Performance.}
    \Description{Two bar graphs comparing sliding window update latency between SIVF and Vanilla Faiss on SIFT1M and GIST1M datasets.}
    \label{fig:eval_sliding}
\end{figure}

As shown in Figure~\ref{fig:eval_sliding}, the standard Faiss baseline suffers from high latency due to the CPU-GPU roundtrip required for index reconstruction, causing system freezes of 355~ms (SIFT1M) and $>$ 1.1~s (GIST1M) per update.
In contrast, SIVF executes updates strictly in VRAM via slab-based management, reducing per-step latency to $\approx$ 2.2~ms (SIFT1M) and 4.2~ms (GIST1M). This yields speedups of $163\times$ and $262\times$, respectively. Notably, the performance gap widens with dimensionality (GIST1M), where the baseline is bottlenecked by PCIe saturation. SIVF eliminates this bottleneck, maintaining single-digit millisecond latency and transforming the GPU IVF index into a real-time streaming engine.

\subsubsection{Long-term streaming}

\begin{table}[t]
\centering
\caption{Tail latency of deletion operations over 1,000 streaming steps (Batch Size $B=1,000$).}
\label{tab:tail_latency}
\begin{tabular}{l|c|cc|c}
\toprule
\textbf{Dataset} & \textbf{Dim} & \textbf{Avg (ms)} & \textbf{P99 (ms)} & \textbf{Max (ms)} \\
\midrule
SIFT1M  & 128 & 0.08 & 0.10 & 0.58 \\
GIST1M  & 960 & 0.08 & 0.10 & 0.53 \\
\bottomrule
\end{tabular}
\end{table}

To validate SIVF's stability under long-term streaming, we conducted a fine-grained benchmark with a reduced batch size of $B=1,000$ over 1,000 continuous steps.
As shown in Table~\ref{tab:tail_latency}, the results confirm two critical characteristics.
First, the latency scales linearly: reducing the batch size by 10$\times$ (compared to the standard batching in Fig.~\ref{fig:eval_deletion}) reduces the average latency proportionally (from $\approx0.89$ ms to $\approx0.08$ ms), confirming the O(1) nature of our bitmap operations.
Second, the performance is exceptionally stable: the 99th percentile (P99) latency is nearly identical to the average (0.10 ms vs 0.08 ms) for both SIFT1M (128D) and GIST1M (960D).
This demonstrate that SIVF's lock-free retry mechanism introduces negligible jitter even under continuous churn, and its performance remains strictly dimension-agnostic.

\subsubsection{Mixed operations}

\begin{table}[t]
\centering
\small
\caption{Search latency stability under mixed workload.}
\label{tab:search_stability}
\begin{tabular}{l|c|cc}
\toprule
Dataset & Dim & Avg (ms) & P99 (ms) \\
\midrule
SIFT1M  & 128 & 0.25 & 0.41 \\
GIST1M  & 960 & 0.69 & 0.78 \\
\bottomrule
\end{tabular}
\end{table}

We evaluated search stability by interleaving query batches within the sliding window loop (Insert $\rightarrow$ Search $\rightarrow$ Delete). 
The results confirm that SIVF maintains sub-millisecond retrieval latency even under continuous mutation. 
On SIFT1M, the 99th percentile (P99) search latency was 0.41 ms (Avg: 0.25 ms). 
Notably, on the high-dimensional GIST1M, the system exhibited negligible jitter with a P99 of 0.78 ms (Avg: 0.69 ms). 
This stability verifies that our slab-based memory management prevents the fragmentation-induced performance degradation typical of dynamic GPU indices.

\subsection{Resource Utilization}
\label{sec:hardware_analysis}

\subsubsection{Computation}

To understand the micro-architectural drivers of the observed performance divergence, we profiled the GPU execution pipeline during the streaming update phase using NVIDIA Nsight Systems and Nsight Compute.
Table~\ref{tab:profiling_breakdown} presents the breakdown of the total execution time.
The baseline implementation reveals a severe resource mismatch: 53.2\% of the runtime is consumed by host-to-device data transfer (\texttt{cudaMemcpyAsync}) via the PCIe bus, and an additional 39.2\% is spent on dynamic memory management overheads (\texttt{cudaMalloc} or \texttt{cudaFree}).
Consequently, the actual GPU compute units (SMs) remain idle for over 96\% of the update cycle, with only 3.2\% of time utilized for kernel execution.
In contrast, SIVF's GPU-resident architecture eliminates these PCIe roundtrips and OS-level allocations.
As a result, the execution profile is transformed: 95.0\% of the time is dedicated to active kernel execution, confirming that SIVF successfully shifts the workload from being I/O-bound to compute-bound.

\begin{table}[t]
\caption{Breakdown of GPU time during streaming updates.}
\label{tab:profiling_breakdown}
\centering
\begin{tabular}{l|r|r}
\toprule
Operation Category & GPU IVF & SIVF (Ours) \\
\midrule
Data Transfer (PCIe) & 53.2\% & $<$ 1.0\% \\
Memory Mgmt (\texttt{malloc}/\texttt{free}) & 39.2\% & $<$ 1.0\% \\
Compute (Active Kernels) & 3.2\% & 95.0\% \\
Other Overheads & 4.4\% & $\approx$ 4.0\% \\
\bottomrule
\end{tabular}
\end{table}

Further profiling with Nsight Compute confirms that SIVF saturates on-device resources.
During the ingestion phase on SIFT1M, the quantization kernels achieve a Streaming Multiprocessor (SM) utilization of 49.0\%, indicating a healthy compute-bound workload.
Simultaneously, the slab selection kernels exhibit an effective memory bandwidth of 150.3 GB/s.
This demonstrates that despite the irregular memory access patterns inherent to linked-slab traversals, SIVF's warp-aligned design ($C=32$) maintains sufficient coalescing to utilize a significant fraction of the device's memory bandwidth, a property unattainable by pointer-based CPU data structures.

\subsubsection{Memory Efficiency and Scalability}
\label{sec:eval_memory}

\begin{figure}[t]
    \centering
    \includegraphics[width=1.0\linewidth]{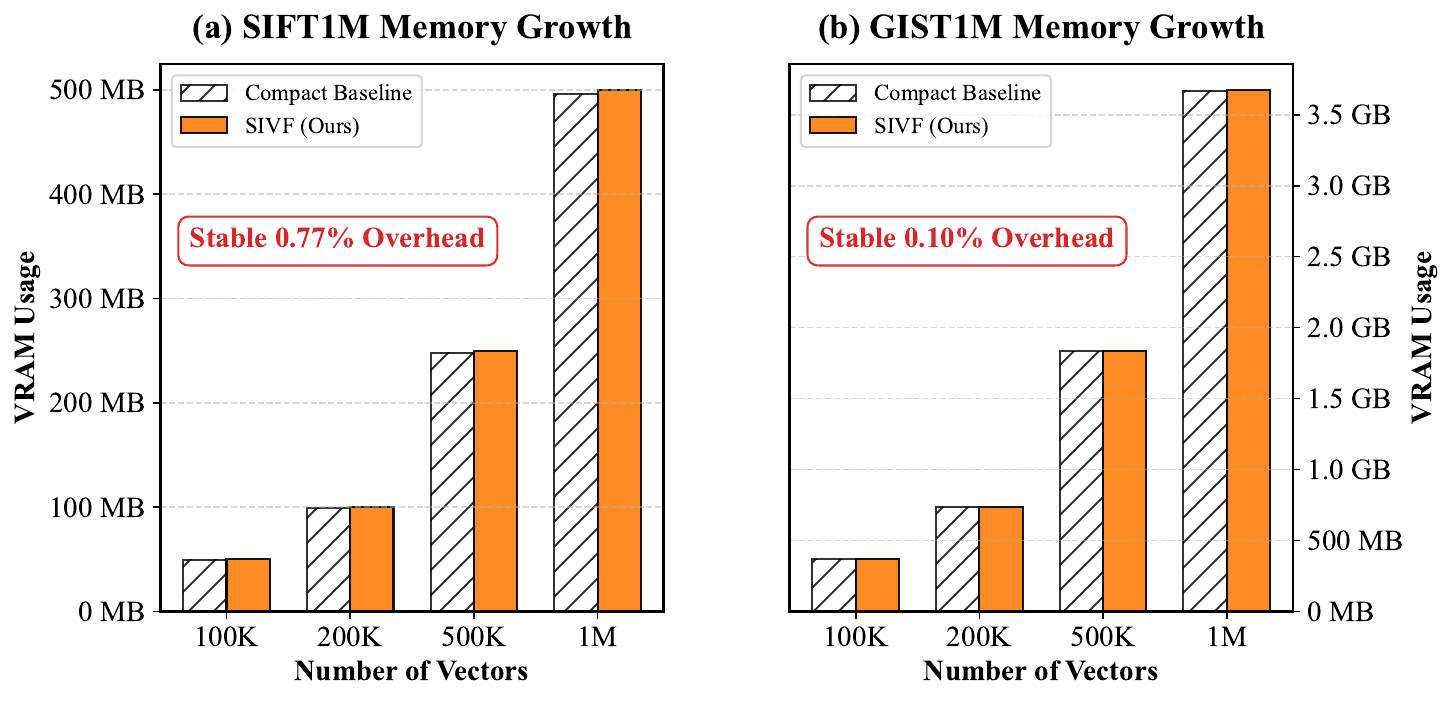}
    \caption{Memory Efficiency and Scalability. 
    }
    \Description{Two line graphs comparing memory usage between SIVF and a compact baseline for SIFT1M and GIST1M datasets, showing SIVF's efficient memory scalability.}
    \label{fig:eval_memory}
\end{figure}

A common concern with dynamic graph-based or list-based indices is the potential for memory bloating due to structural overhead, such as pointers and metadata, as well as fragmentation. To quantify the space complexity of SIVF, we conducted a memory footprint analysis comparing the allocated VRAM usage of SIVF against a theoretically compact array baseline across varying dataset sizes ranging from 100K to 1M vectors.
Figure~\ref{fig:eval_memory} illustrates the memory growth trends for both SIFT1M and GIST1M. The results demonstrate that SIVF exhibits deterministic linear scalability with negligible structural overhead. For SIFT1M ($d=128$), the overhead stabilizes at 0.77\%, while for the high-dimensional GIST1M ($d=960$), it is merely 0.10\%. This efficiency stems from our coarse-grained slab design, where the metadata cost (128-byte header) is amortized over 32 vectors. 

We verified the efficacy of our memory reclamation strategy. In a stress test (not shown in the figure) involving the deletion of 50\% of the dataset followed by immediate re-insertion, SIVF maintained a constant memory footprint without triggering OS-level deallocation. The deletion operation completed in sub-millisecond time per batch (e.g., 4.04~ms for 100K GIST vectors), confirming that memory slots are logically reclaimed via bitmap toggling and immediately available for reuse. This zero-cost reclamation ensures that SIVF can sustain long-running streaming workloads without suffering from memory leaks or fragmentation-induced bloat.

\subsection{Non-IVF Indexes for Streaming Vectors}
\label{sec:eval_non_ivf}

To position SIVF within the broader indexing landscape, we benchmarked it against five representative non-IVF architectures: GPU CAGRA~\cite{10597683} (graphs), Faiss GPU Flat (brute-force baseline, i.e., no index), HNSW~\cite{malkov2018efficient} (dynamic graphs), (3) LSH~\cite{10.1145/997817.997857} (legacy hash-based approaches), and (4) NSG~\cite{10.14778/3303753.3303754} (static graphs). Table~\ref{tab:landscape_full} summarizes ingestion throughput and deletion latency across three datasets: SIFT1M, T2I-1B, and GIST1M.

\begin{table}[t]
\caption{Add throughput (K vec/s) and delete latency (ms).}
\label{tab:landscape_full}
\centering
\resizebox{\linewidth}{!}{
\begin{tabular}{l|rr|rr|rr}
\toprule
\multirow{2}{*}{\textbf{Method}} & \multicolumn{2}{c|}{\textbf{SIFT1M}} & \multicolumn{2}{c|}{\textbf{T2I-1B}} & \multicolumn{2}{c}{\textbf{GIST1M}} \\
 & Add & Delete & Add & Delete & Add & Delete \\
\midrule
Faiss Flat~\cite{douze2024faiss} & 9,308 & 833 & 5,957 & 1,163 & 1,907 & 1,040 \\
nuVS CAGRA~\cite{10597683} & 305 & 3,030 & 259 & 3,773 & 97.1 & 10,215 \\
HNSW~\cite{malkov2018efficient} & 25.1 & 39,835 & 25.8 & 39,141 & 0.6 & 334,113 \\
NSG~\cite{10.14778/3303753.3303754} & 3.7 & 266,224 & 3.2 & 314,315 & 0.7 & 278,101 \\
LSH~\cite{10.1145/997817.997857} & 823 & 13.7 & 461 & 16.4 & 38.4 & 8.5 \\
\textbf{SIVF} (this work) & \textbf{2,229} & \textbf{0.95} & \textbf{5,183} & \textbf{1.52} & \textbf{452} & \textbf{1.31} \\
\bottomrule
\end{tabular}
}
\end{table}

\begin{figure*}[t!]
  \centering
  \includegraphics[width=\linewidth]{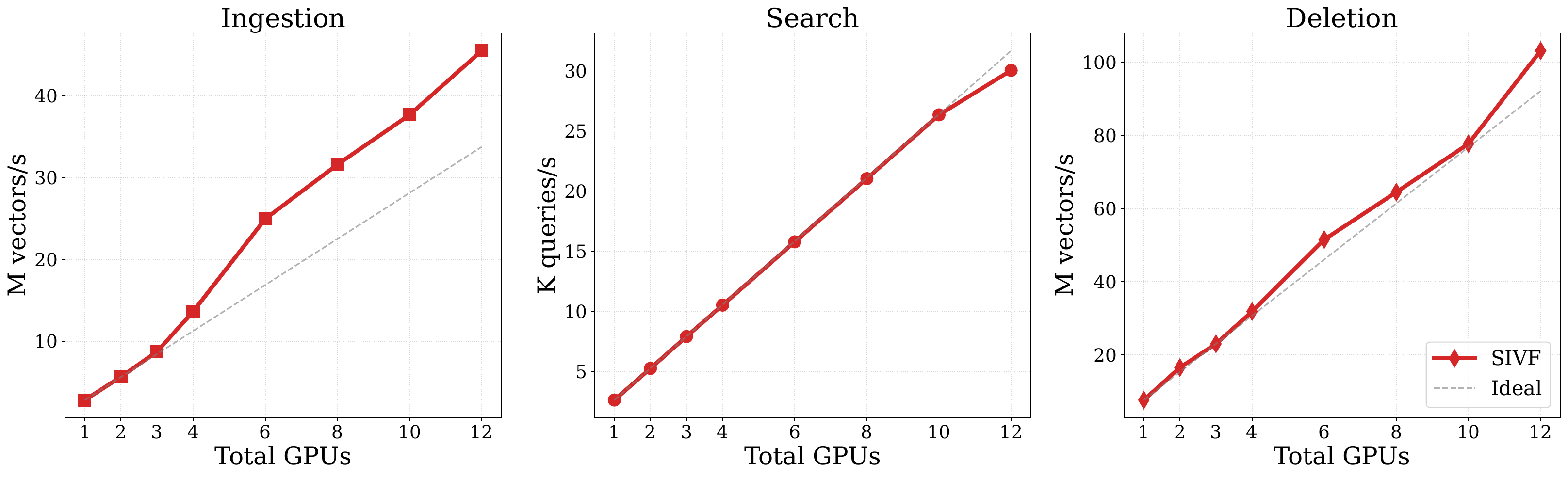}
  \caption{Scalability across a 4-node cluster of 12 GPUs.}
  \Description{Three line graphs showing the scalability of SIVF core operations (ingestion, search, deletion) across a four node cluster with up to 12 GPUs.}
  \label{fig:scalability}
\end{figure*}

\begin{figure}[t!]
  \centering
  \includegraphics[width=\linewidth]{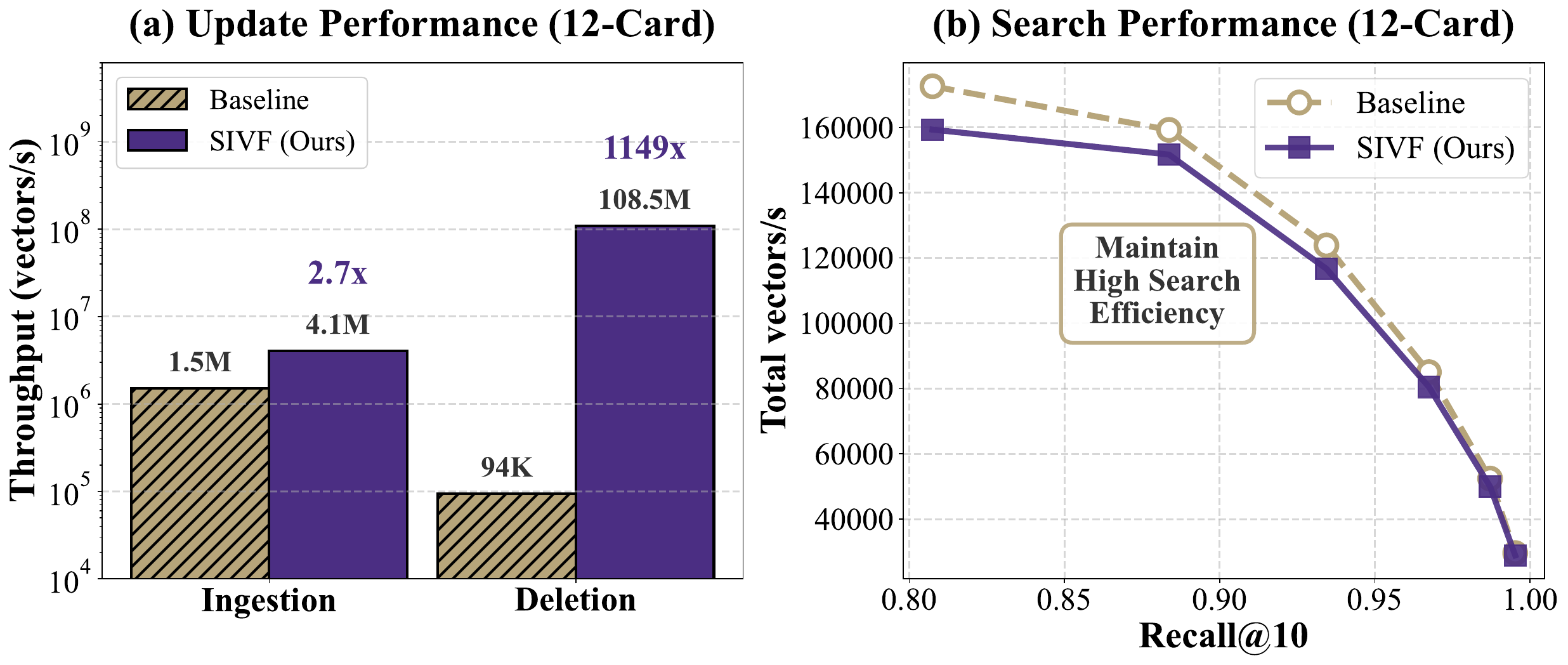}
  \caption{Distributed performance on the 12-GPU cluster (DINO10B dataset).}
  \Description{Two side-by-side charts on DINO 1024d. Left: Bar chart showing 2.69x ingestion speedup and 1150x deletion speedup (log scale). Right: Pareto frontier showing SIVF maintains high search efficiency (approx. 160K vectors/s) across 12 GPUs.}
  \label{fig:dino_perf_12gpu}
\end{figure}

Table~\ref{tab:landscape_full} reports the results of vector ingestion and deletion in a streaming context across three datasets.
CPU-based graph indices (HNSW, NSG) suffer catastrophic performance degradation on high-dimensional data, dropping to merely 600 vec/s on GIST1M due to cache thrashing. They lack native support for deletion, necessitating full index reconstruction that incurs prohibitive latencies (e.g., over 330 seconds for HNSW on GIST1M).
While GPU Flat achieves high ingestion throughput by bypassing indexing overhead, its deletion latencies remain high ($>$ 1s) due to $O(N)$ data compaction and CPU-GPU synchronization.
Notably, the state-of-the-art GPU graph index, CAGRA, exhibits severe limitations in streaming scenarios; enforcing true physical deletion to prevent memory leaks necessitates reconstruction, causing latencies to spike to over 10 seconds.
In contrast, SIVF emerges as the only architecture enabling holistic stream processing: it sustains GPU-class ingestion rates (e.g., 452K vec/s on GIST1M, $4.6\times$ faster than CAGRA) while providing millisecond eviction with immediate slot reuse (via bitmap toggling), and slab-level reclamation when a slab becomes empty. 

\subsection{Scalability on Multi-Node GPU Clusters}
\label{sec:eval_multigpu}

We evaluated the scalability performance of SIVF on the Chameleon TACC site~\cite{keahey2020lessons}. 
Each node is equipped with dual-socket Intel Xeon E5 CPUs running at 2.00 GHz and 128 GB of RAM. 
The compute power is provided by NVIDIA Tesla P100 GPUs.
The experimental environment leverages a total of 12 GPUs distributed across 4 physical nodes (4 GPUs each on \texttt{c11-01} and \texttt{c11-04}, and 2 GPUs each on \texttt{c11-03} and \texttt{c11-22}).

Figure~\ref{fig:scalability} shows near-linear scalability for 128-dimensional vectors. On 12 GPUs, ingestion peaks at 45.5 million vec/s, and search reaches 30.1K vectors/s (95\% efficiency relative to the single-GPU baseline). Deletion exhibits super-linear scaling, achieving 103.2 million vectors/s. This efficiency boost stems from the compute-bound nature of the workload: expanding the cluster increases aggregate GPU cache capacity and reduces per-device contention for bitmap synchronization, thereby outperforming linear projections.

We evaluate SIVF against the primary baseline (Faiss GPU IVFFlat) using the 1024-dimensional DINO10B dataset~\cite{9709990} on the 12-GPU cluster.
Figure~\ref{fig:dino_perf_12gpu}(a) highlights SIVF's efficiency in handling data updates. For ingestion, SIVF achieves a 2.69$\times$ speedup over the baseline (4.07M vectors/s vs. 1.51M vectors/s).
The most significant performance gap is observed in deletion operations. SIVF achieves a 1150$\times$ speedup, processing over 108M deletions per second compared to the baseline's 94K. 
Figure~\ref{fig:dino_perf_12gpu}(b) plots the Pareto frontier of Recall@10 versus throughput. SIVF exhibits zero recall loss and retains over 95\% of the baseline's throughput.

\section{Conclusion}
\label{sec:conclusion}

This work addresses the immutability issues in GPU-accelerated Inverted File (IVF) indices caused by static memory layouts. We present SIVF, a GPU-native index enabling high-velocity, in-place updates via slab-based memory management. By leveraging a lock-free update mechanism and optimized slab traversal, SIVF effectively masks the synchronization overhead typical of dynamic indexing, allowing the system to fully exploit hardware parallelism. Evaluation on a 12-GPU cluster demonstrates that SIVF achieves near perfect linear scalability, reaching 4.07M vectors/s for ingestion and 108.5M vectors/s for deletion. Furthermore, SIVF maintains high-accuracy search performance at 159.3K queries/s while keeping memory overhead negligible.

Despite these advances, SIVF currently operates strictly within GPU VRAM, which bounds the maximum dataset size to the aggregate device memory. Future work will explore hierarchical storage architectures, transparently offloading cold slabs to host memory or NVMe storage to support billion-scale sliding-window sizes. Additionally, while the current slab reclamation strategy effectively recycles empty nodes, we plan to investigate adaptive background compaction policies to further reduce internal fragmentation under highly skewed deletion patterns. 


\section*{Acknowledgment}
Results presented in this paper were obtained using the Chameleon testbed supported by the National Science Foundation.
We thank Meta for merging some of our code into the main Faiss branch in March 2026.

\bibliography{ref,ann,gpu}
\bibliographystyle{acm}

\clearpage
\appendix

\section{Appendix}

\subsection{Full Proofs}

\begin{theorem}[Parallel ingestion is safe and linearizable]
\label{thm:ingestion}
Consider any concurrent execution of Algorithm~\ref{alg:sivf_ingestion}.
For every insertion that returns successfully, there exists a unique slot $(s,o)$ such that:
(i) the insertion operation writes the payload and id into that slot, then sets bit $o$ in \textup{\texttt{validity\_bitmap}} exactly once,
(ii) once the bit is set, the slot contains a fully initialized payload and id, and
(iii) the insertion can be linearized at the atomic bit set operation \textup{\texttt{atomicOr}} that makes the slot visible.
\end{theorem}

\begin{proof}
We first argue the uniqueness of the chosen slot for a successful insertion.

In the existing head case, the code reads 
\[ c=\texttt{slab\_meta}[h].\texttt{valid\_count}, \] 
and attempts to reserve index $c$ using an atomic compare-and-swap (CAS). 
Since the CAS operation atomically verifies that the count is still $c$ before incrementing it to $c+1$, only one thread can succeed for any specific snapshot of the counter.
Therefore, among concurrent contenders for the same slab $h$, at most one insertion obtains a given index $c$. 
In the new slab case, the thread publishes its new slab as the head using \texttt{atomicCAS(\&H[$\ell$], h, s\_new)}; only one thread can win this publication for a fixed old head value $h$,
and the winner uses slot $o=0$ in its newly allocated slab. Thus, every successful insertion selects exactly one slot $(s,o)$.

We next argue that the chosen slot is initialized before it becomes visible.
In both code paths, after reserving the slot, the writer performs the payload write to \texttt{slab\_data[$s$][$o$]} and
the id write to \texttt{slab\_ids[$s$][$o$]}, then writes $\mathcal{T}(v_{id})\gets\langle s,o\rangle$, executes
\verb|__threadfence()|, and only after the fence performs \texttt{atomicOr} to set bit $o$ in the bitmap.
The memory consistency semantics of \verb|__threadfence()| ensure that all prior global writes are committed to memory before any subsequent atomic operation by the same thread becomes visible to other threads.
Therefore, any observer that sees bit $o$ as set is guaranteed to see the fully initialized payload and id.

Finally, we justify linearizability of a successful insertion with the atomic bit set as the linearization point.
The membership predicate used throughout the design is exactly the bitmap bit. Before the \texttt{atomicOr}, the bit is 0,
so the slot is logically absent to any search warp. After the \texttt{atomicOr} takes effect, the bit is 1, so the slot is logically present.
Since \texttt{atomicOr} executes atomically on the bitmap word, the state transition $0 \to 1$ occurs at a single instant.
At that instant, the slot becomes visible and, by the fence ordering established above, it is valid. Thus, the insertion is linearizable.
\end{proof}

\begin{theorem}[Search safety under concurrent ingestion and deletion]
\label{thm:search}
In any concurrent execution of Algorithms~\ref{alg:sivf_ingestion}, \ref{alg:sivf_search}, and \ref{alg:sivf_delete},
every payload and id read performed by Algorithm~\ref{alg:sivf_search} is fully initialized.
\end{theorem}

\begin{proof}
Algorithm~\ref{alg:sivf_search} reads a slot $(s,j)$ only if it first observes that bit $j$ in
\texttt{slab\_meta[$s$].validity\_bitmap} is set. When the search observes the bit set, the slot must have been published
by a successful insertion. By Theorem~\ref{thm:ingestion}, the publication point is the atomic bit set in
Algorithm~\ref{alg:sivf_ingestion}, and the payload and id writes occur before the fence, which ensures they complete before the bit set.
The same memory consistency guarantee implies that once the bit set is visible to a search warp, the earlier payload and
id writes are also visible. Therefore, the search cannot observe a partially initialized payload or id.

Deletion does not write payloads or ids. Algorithm~\ref{alg:sivf_delete} only clears the validity bit and, upon slab reclamation, 
resets the bitmap to 0 before returning the slab to the free pool. 
Hence, even if a search races with a delete or a subsequent slab reuse, the primary guard is the bitmap state.
If the search sees the bit as 0, it skips the slot. If it sees the bit as 1, it reads a payload and id that were fully initialized
before the corresponding bitmap set operation (whether from the original insertion or a reuse). 
In neither case does the search read corrupted or partially written data.

Algorithm~\ref{alg:sivf_search} also terminates because its traversal is bounded, and it includes an explicit self-loop guard
(\texttt{md.next == s}) before following \texttt{next}. Concurrent insertions can add a new head slab, but they do not require the
search to revisit already visited slabs within the traversal bound.
\end{proof}

\begin{theorem}[Lazy eviction is linearizable and makes deleted vectors invisible]
\label{thm:delete}
Consider any concurrent execution of Algorithms~\ref{alg:sivf_search} and \ref{alg:sivf_delete}.
For any delete request on $u$ such that 
$\mathcal{T}[u] \neq \textnormal{\texttt{INVALID}}$
and decodes to
$(\text{idx}_{slab},\text{idx}_{slot})$, define
\[
mask = \mathbin{\sim}(1 \ll \text{idx}_{slot}).
\]
Then the atomic operation
$old\_map \gets atomicAnd()$ 
in Algorithm~\ref{alg:sivf_delete} constitutes the linearization point for logical deletion. After this atomic operation takes effect,
the slot becomes logically absent from the search space. Repeated deletes of the same id are idempotent and safe.
\end{theorem}

\begin{proof}
If $\mathcal{T}[u]=\texttt{INVALID}$, Algorithm~\ref{alg:sivf_delete} performs no state change and the delete is linearized
at the \texttt{INVALID} check.

Otherwise, the delete decodes \texttt{coord} into $(\text{idx}_{slab},\text{idx}_{slot})$, computes
$mask$, and executes
\[old\_map \gets \texttt{atomicAnd}(\text{slab\_meta}[\text{idx}_{slab}].\text{bitmap}, mask)\] to clear the corresponding bit.
Since \texttt{atomicAnd} is an atomic read-modify-write instruction on the GPU, the transition of the validity bit from 1 to 0 (or 0 to 0) occurs at a single, indivisible instant.
We designate this instant as the linearization point.

Algorithm~\ref{alg:sivf_search} uses the validity bit as the sole membership predicate.
Because the search warp reads the bitmap atomically before accessing any payload, any search that observes the bitmap
after the linearization point will see the bit as 0, skip the slot, and exclude the vector from the results.
Physical payload reclamation is not required for correctness because logical membership is fully determined by the validity bit.

To handle duplicates, Algorithm~\ref{alg:sivf_delete} uses the return value \texttt{old\_cnt} to detect if it was the specific thread
that caused the $1 \to 0$ transition. If so, it performs bookkeeping (e.g., updating counters). If the bit was already 0,
the operation is effectively a no-op on the logical state. Thus, the deletion mechanism is idempotent.
\end{proof}

\end{document}